\documentclass{article}
\usepackage{epsfig}
\usepackage[hang,nooneline]{subfigure}
\usepackage{amsmath}
\usepackage{amsfonts}
\usepackage{amssymb}
\usepackage{graphicx}
\usepackage{epstopdf}
\usepackage{amsmath}
\usepackage{graphicx} \usepackage{float}
\usepackage{amsthm}
\usepackage{comment}
\usepackage{color}

\def\mathcolour#1#{\mathcoloraux{#1}}
\newcommand*{\mathcoloraux}[3]{%
  \protect\leavevmode
  \begingroup
    \color#1{#2}#3%
  \endgroup
}

\usepackage[svgnames]{xcolor}

\usepackage{xcolor}

\newenvironment{myitemize}{
\begin{itemize}
 \setlength{\itemsep}{1pt}
 \setlength{\parskip}{0pt}
 \setlength{\parsep}{0pt}}{\end{itemize}
}

\newenvironment{myenumerate}{
\begin{enumerate}
 \setlength{\itemsep}{1pt}
 \setlength{\parskip}{0pt}
 \setlength{\parsep}{0pt}}{\end{enumerate}
}

\begin{document}

\title{\vspace{-2cm}Rule Primality, Minimal Generating Sets, Turing-Universality and Causal Decomposition in Elementary Cellular Automata}
\author{
{\bf J\"urgen Riedel\footnote{jurgen.riedel@labores.eu}} \\ Algorithmic Nature Group, LABORES, Paris, France; and\\
Algorithmic Dynamics Lab, Centre for Molecular\\Medicine, Karolinska Institute\\\\ and \\\\
{\bf Hector Zenil\footnote{hector.zenil@algorithmicnaturelab.org}}\\Information Dynamics Lab, Unit of Computational Medicine,\\ SciLifeLab, Centre for Molecular Medicine, Department\\ of Medicine Solna, Karolinska Institute;\\
Department of Computer Science, University of Oxford, UK; and\\
Algorithmic Nature Group, LABORES, Paris, France.
}
\date{}

\newtheorem{mytheorem}{Theorem}
\newtheorem{mydefinition}{Definition}
\newtheorem{myobservation}{Observation}
\newtheorem{myconjecture}{Conjecture}

\maketitle

\bigskip

\begin{abstract}
We introduce several concepts such as \textit{prime} and \textit{composite rule}, tools and methods for causal composition and decomposition. We discover and prove new universality results in ECA, namely, that the Boolean composition of ECA rules 51 and 118, and 170, 15 and 118 can emulate ECA rule 110 and are thus Turing-universal coupled systems. We construct the 4-colour Turing-universal cellular automaton that carries the Boolean composition of the 2 and 3 ECA rules emulating ECA rule 110 under multi-scale coarse-graining. We find that rules generating the ECA rulespace by Boolean composition are of low complexity and comprise \textit{prime rules} implementing basic operations that when composed enable complex behaviour. We also found a candidate minimal set with only 38 ECA \textit{prime rules}---and several other small sets---capable of generating all other (non-trivially symmetric) 88 ECA rules under Boolean composition.\\

\end{abstract}

\section{Notation and Preliminaries}

The following definitions follow the notation in~\cite{Powley2009}. A {\it cellular automaton} (CA) is a tuple $\langle S, (\mathbb{L}, +), T, f \rangle$  with a set $S$ of states, a lattice $\mathbb{L}$ with a binary operation $+$, a neighbourhood template $T$, and a local rule $f$.

The {\it set of states} $S$ is a finite set with elements $s$ taken from a finite alphabet $\Sigma$ with at least two elements. It is common to take an alphabet composed entirely of integers modulo $s$: $\Sigma = \mathbb{Z}_s = \{0,...,s-1\}$. 
An element of the lattice $i\in \mathbb{L}$ is called a cell. The lattice $\mathbb{L}$ can have $D$ dimensions and can be either infinite or finite with cyclic boundary conditions.

The {\it neighbourhood template} $T=\langle\eta_1,...,\eta_m\rangle$ is a sequence of $\mathbb{L}$. In particular, the neighbourhood of cell $i$ is given by adding the cell $i$ to each element of the template $T$: $T=\langle i+\eta_1,...,i+\eta_m\rangle$. 
Each cell $i$ of the CA is in a particular state $c[i] \in S$. A {\it configuration} of the CA is a function $c: \mathbb{L} \rightarrow S$. The {\it set of all possible configurations} of the CA is defined as $S_\mathbb{L}$.

The {\it evolution of the CA} occurs in discrete time steps $t=0,1,2,...,n$. The transition from a configuration $c_t$ at time $t$ to the configuration $c_{(t+1)}$ at time $t+1$ is induced by applying the local rule $f$. The local rule is to be taken as a function $f: S^{|T|} \rightarrow S$ which maps the states of the neighbourhood cells of time step $t$ in the neighbourhood template $T$ to cell states of the configuration at time step $t+1$:
\begin{equation}
c_{t+1}[i]=f\left(c_t[i+\eta_1],...,c_t[i+\eta_m]\right )
\end{equation}
The general transition from configuration to configuration is called the {\it global map} and is defined as: $F: S^\mathbb{L} \rightarrow S^\mathbb{L}$.

In the following we will consider only 1-dimensional (1-D) CA as introduced by Wolfram~\cite{StephenWolfram1983,nks}. The lattice can be either finite, i.e. $\mathbb{Z}_N$, having the length $N$, or infinite, $\mathbb{Z}$. In the 1-D case it is common to introduce the {\it radius} of the neighbourhood template which can be written as $\langle -r,-r+1,...,r-1,r \rangle$ and has length $2 r+1$ cells. With a given radius $r$ the local rule is a function $f: \mathbb{Z}_{|S|}^{{|S|}^{(2r+1)}} \rightarrow \mathbb{Z}_{|S|}$ with $\mathbb{Z}_{|S|}^{{|S|}^{(2r+1)}}$ rules. The so called Elementary Cellular Automata (ECA) with radius $r=1$ have the neighbourhood template $\langle -1,0,1\rangle$, meaning that their neighbourhoods comprise a central cell, one cell to the left of it and one to the right. The rulespace for ECA contains $2^{2^{3}}=256$ rules. Here we consider non-equivalent rules subject to the operations complementation, reflection, conjugation and joint transformation (combining reflection and conjugation) (see Supplementary Information). For example, the number of reduced rules for ECA is 88 (see Supplementary Information).

In order to keep the notation simple, we adopt the following definitions \cite{NavotGoldenfeld2006}. A cellular automaton at time step $t$ $A=(a(t),\{S_{A} \},f_{A})$ is composed of a lattice $a(t)$ of cells that can each assume a value from a finite alphabet ${S_{A}}$. A single cell is referenced as $a_{n}(t)$. The update rule $f_{A}$ for each time step is defined as $f_{A}: \{S^{2^{2 r+1}}\} \rightarrow \{S_{A}\}$ with $a_{n}(t+1)=f_{A}[a_{n-1}(t),a_{n}(t),a_{n+1}(t)]$. The entire lattice gets updated through the operation $f_{A}  a(t)$.

\subsection{CA Typical Behaviour and Wolfram's Classes}

Wolfram also introduced~\cite{nks} an heuristic for classifying computer programs by inspecting the behaviour of their space-time diagrams. Computer programs behave differently for different inputs. It is possible, and not uncommon, however, to analyze the behaviour of a program asymptotically according to an initial condition metric~\cite{zenilca,zenilchaos}.

Wolfram's classes can be characterized as follows:

\begin{myitemize}
\item Class 1. Symbolic systems which rapidly converge to a uniform state. Examples are rules 0, 32 and 160.
\item Class 2. Symbolic systems which rapidly converge to a repetitive or stable state. Examples are rules 4,
108 and 218.
\item Class 3. Symbolic systems which appear to remain in a random state. Examples are rules 22, 30, 126
and 189.
\item Class 4. Symbolic systems which form areas of repetitive or stable states, but which also form structures
that interact with each other in complicated ways. Examples are rules 54 and 110.
\end{myitemize}

We use the concept of A Wolfram class as a guiding index that popularly assigns some typical behaviour to every ECA even though we have also shown that such distinction is not of fundamental nature~\cite{riedelzenil}. Here, however, it will be useful to study this idea of typical behaviour of rules that are capable of emulating others when included in minimal emulation sets.

In~\cite{cook,nks}, it was shown that at least one ECA rule can perform Turing universal computation. It is still an open question whether other rules are capable of Turing universality, but some evidence suggests that they may be, and that cellular automata rules and random computer programs are, in general, highly programmable and candidates of computation universality~\cite{riedelzenil}.

Here we extend results reported in~\cite{StephenWolfram1986} regarding the Boolean composition of ECA. We introduce minimal generating sets as candidates for being able to generate all ECA rules, and we introduce new Turing-universality results in ECA by composition and an associated non-ECA Turing-universal CA implementing the composition.

\section{Methods}

\subsection{Rule Composition}

Rule composition for a pair of CA, i.e. \textrm{rule $C$} = $\textrm{rule A} \circ \textrm{rule B}$, is defined as $f^{1}_{C}a(0)=f^{1}_{B} \circ (f^{1}_{A} a(0))$. The lattice output of rule $A$ is the input of rule $B$. One can say that the rule composition of rule $A$ and rule $B$ yields rule $C$. Rule $C$ can be composed out of rule $A$ and rule $B$. The whole evolution of the composite $\textrm{rule A} \circ \textrm{rule B}$ is $f_{B}  \circ f_{A} a(2t) =f_{C} a(t)$ which is 
as long as the whole evolution of $\textrm{rule C}$. More generally one can compose rule $A$ out of n rules $A_n$: $f^{1}_{A}a(0)=f^{1}_{A_1} \circ f^{1}_{A_2}  \ldots \circ f^{1}_{A_n} a(0)$. The whole evolution is $f_{C} a(t)=f_{A_1} \circ f_{A_2} \ldots \circ f_{A_n}  a(n t)$.

In order to find the CA in a higher rule space that implement the Boolean composition of CA in lower rule spaces (e.g. ECA) we introduce the concept of causal separability.

\begin{mydefinition}
A space time evolution of a function $C:S\rightarrow S^\prime$ (e.g. a CA)  is minimally causally separable if, and only if, the rule icon network (see Fig.~\ref{fig_map_50_37b}) of the rule $R$ of $C$ is the smallest rulespace in which $R$ is separable into $|S|$ disconnected networks.
\end{mydefinition}

Fig.~\ref{fig_map_50_37b} illustrates the basics of (non-)separability under a rule composition of ECA rules, an example demonstrating that the ECA rulespace is not closed under Boolean composition. This will help us find the CA rule in a higher space that implements the emulation of rule 110 under composition of ECA rules (see Fig.~\ref{fig_4color_map_110} in the Supplementary Material).

\subsection{Minimal Generating ECA Rule Sets}

The questions driving our experiments thus led us to the problem of finding the minimal rule set that generates the full ECA space.

The implementation of the algorithm, and thus checking for rule $N$-primality, is equivalent to the formal equivalence checking process in electronic design automation to formally prove that two representations of a circuit design exhibit exactly the same behaviour, which is 
reducible to the Boolean satisfiability problem (SAT). Since the SAT problem is NP-complete, only algorithms with exponential worst-case complexity can carry this test.

We therefore proceeded by sampling methods. One strategy is to start from a subset of ECA rules, composing and finding the emulations that fall back into the ECA rulespace, because only a subset of all possible rule pairs of a given rulespace lead to a rule which itself is a member of the same rulespace, thus clearly indicating that the ECA under composition is not an algebraic structure, as it is not closed under composition. If a rule composition remains in the same rulespace, the rule tuples, after application of the successive rules of the rule composition, map to a cell state which is one-to-one (see Fig.~\ref{fig_map_50_37}). 
However, this mapping is not one-to-one all the time and the resulting rule composition leaves the rulespace of the constituent rules (again Fig. \ref{fig_map_50_37}). In this paper we will focus on both cases. We investigate the former case where the composite rule remains in the same rulespace, and set our focus on Turing-universality. Another strategy is to start from a subset of ECA rules and start finding the pair of rules that can emulate the rules in question, then move to another rule, knowing the compositions that were already found in the previous steps. However, all sampling approaches have their own potential problems because the result may be dependant on the initial subset of rules chosen to start the exploration. So we also employed a procedure akin to bootstrapping, where we re-sampled the ECA rulespace with different seeds (initial ECA rules) to infer the compositions from a new sample of the same rulespace and sought out convergence patterns.

We also explore the mapping to a higher colour rulespace in order to analyze the full richness of the rulespace induced by rule composition.

\subsection{Exploratory Algorithms}
\label{samplingalg}

\subsubsection{Sampling algorithm 1}

\begin{figure}[ht!]
\centering
\begin{tabular}{c}
  \label{fig_rule_pair_1}\includegraphics[width=80mm]{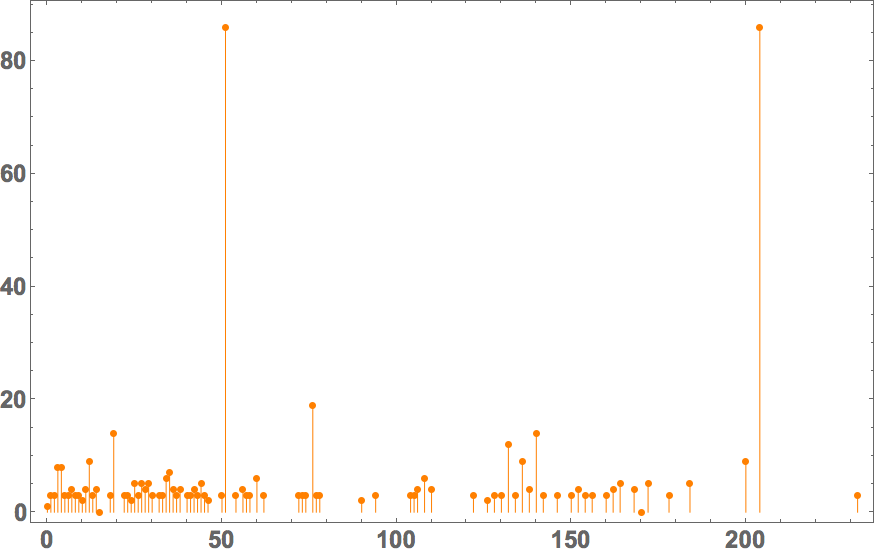}\\
  (a) Distinct count ($y$ axis) of ECA rules which can be emulated\\ by other ECA rules ($x$ axis).
\\[6pt]
 \label{fig_rule_pair_2}\includegraphics[width=80mm]{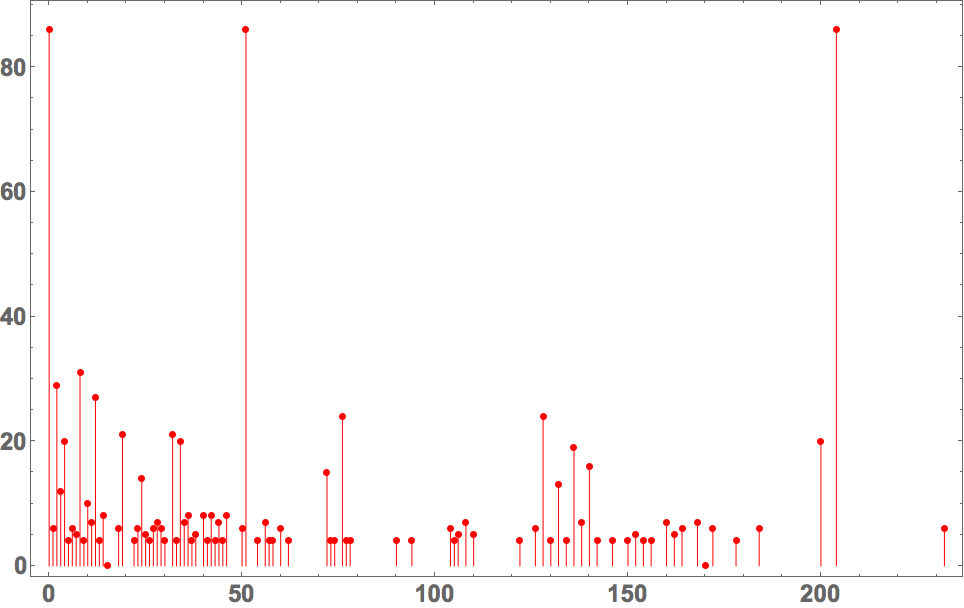}\\ 
 (b) Distinct count ($y$ axis) of ECA rules which form pairs\\with other ECA rules ($x$ axis).
 
 \\[6pt]
\end{tabular}
\caption{\label{fig_rule_pair} (a) Showing the non-uniform distribution count ($y$ axis) of ECA rules which can be emulated by ECA rules ($x$ axis) pairing with another ECA rule. For example, rule pairs containing rule 51 or rule 204 can emulated the most ECA rules. 
(b) Showing the distinct count ($y$ axis) of ECA rules which form pair with ECA rules  ($x$ axis). For example rules 0, 51, and 204 form the most distinct pairs with other ECA rules.
}
\end{figure}

The sampling algorithm 1 is based on the frequency of rule 51 in all rule pairs (i.e. 2-tuple) emulating another rule. On one hand, the number of distinct rules composed of rule 51 is the highest (see Fig.~\ref{fig_rule_pair_1} (a)) (besides rule 204, which is the identity rule). On the other hand, the number of distinct rule forming a composition pair is the highest as well (see Fig.~\ref{fig_rule_pair_2} (b)). Here rules 204  the identity rule, and 0, the annihilator rule, have the same number of distinct rules and are not of importance in this context.

The algorithm searches alongside all rule pairs containing rule 51. If no rule pair containing rule 51 is found, the algorithm searches for other valid rule pairs. It tries to substitute (fold) each rule pair with other rule pairs found in order to form a reduced $n$-tuples. The goal of the algorithm is to minimize set of rule tuples. 

Since rule 51 can be composed of rules 15 and 170, the algorithm substitutes always rule 51 with rule pair (15,170). 

The pseudo-code for the first sampling algorithm is as follows:

\begin{myenumerate}
\item create empty sets prime.rules and ensemble.tuples as well as the set all.ECA.rules containing all 88 ECA rules. 
\item do 
\begin{myenumerate}
    \item Draw a set containing rule tuples which each having composing rules different than the composed rules and the set as a whole containing all 88 ECA rules. This set is called tuple.pool
	\item select all tuples from tuple.pool the composing rules of which are not in the set prime.rules \label{alg1_ref1}
	\item if selection (from step \ref{alg1_ref1}) $=$ empty then break
	\item draw(one random sample) from tuple set
	\item add the composing rule to the set prime.rules
	\item add whole tuple to set ensemble.tuples
     \end{myenumerate}

\item for test.tuples in ensemble.tuples
    \begin{myenumerate}
	\item extract all distinct composing rules from set and put them into prime.rules
	do
	 \begin{myenumerate}
		\item select at random a rule prime.rule from the set test.tuples not in $\{15, 51, 170\}$.
		\item set valid.tuples $=$ (get all tuples which have prime.rule as the composite rule)
		\item remove all tuples from set valid.tuples which have composing rules in set prime.rules
		test.tuples $=$ (in all tuples of set test.tuples replace the composed rule with the composing rules of set valid.tuples)
		\item remove prime.rule from set prime.rules
		\item remove all tuples from test.tuples which have repeating rules.
		\item break if order of set prime.rules $=$ 0
		 \end{myenumerate}
 \end{myenumerate}
	\item prime.rules $=$ extract all distinct composing tuples from set test.tuples
	\item prime.rules $=$ prime.rules $+$ rules not contained in prime.tuples from all.ECA.rules
	\item comp.tuples $=$ test.tuples $+$ \\
	(for rule in primes.rules \\
	      \-\hspace{1cm} tuples $=$ tuples $+$ $\{$rule,\_,\_$\}$ 
	    \\ end for).

 \end{myenumerate}

\subsubsection{Sampling algorithm 2}

This algorithm does not rely on special insight in the of rule composition pairs for ECA rules as it is only relying on random sampling of the whole set of rule composition pairs. \\

Pseudo-code for the second sampling algorithm is as follows:

 \begin{myenumerate}
 \item  select all tuples from set all.tuples which do not have repeating rules
  \item set prime.pool $=$ all 88 ECA rules
  \item initialize as empty sets all.tuples and new.tuples $=$ $\{$ $\}$
  \item initialize as empty set
  selected.rules $=$ $\{$ $\}$
  \begin{myenumerate}
   \item set rule $=$ 
    (pick a random rule from prime.pool excluding selected.rules)
   \item set selected.rules $=$ selected.rules $+$ rule
   \item initialize new.primes, new.rules and rules as empty sets
   \item rules $=$ $\{$rule$\}$ (start with set containing one rule)
   \begin{myenumerate}
    \item tuples $=$ (draw for each rule in rules a tuple from prime.pool which composes rule) 
	  \item check if set tuples only contains composing rules which are not already composites
      \item If length(tuples)$=$0, break
      \item set rules $=$ (all composed rules in set tuples)
      \item set new.rules $=$ newrules $+$ tuples
    \item  else break at 100 steps
     \end{myenumerate}
   	\item new.tuple $=$ fold (definition below) set new.rules to form one tuple adhering to causal order
    \item primes $=$ select composing rules of new.tuple
    \item set new.tuples $=$ new.tuples $+$ new.tuple
    \item if number of distinct rules in set new.tuple $=$ 88 then break
    \item else break at max steps
 \end{myenumerate}
   \item all.tuples $=$ all.tuples $+$ new.tuples
 \item stop at maximal number of trials max
 \end{myenumerate}

The folding function in this algorithm refers to the substitution of composite rules previously found to be emulated by other prime rules in previous iterations of the same the algorithm, i.e. the substitution of rules that can be decomposed in other prime rules.

\subsection{Primality and Rule (De)Composition}

\begin{figure}
\centering
\begin{tabular}{c}
  \label{fig_map_54}\includegraphics[width=120mm]{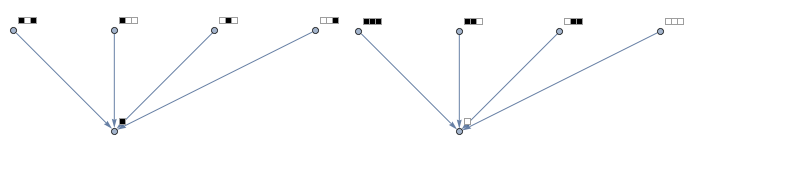}\\
  (a) Rule mapping ECA rule 54.
\\[6pt]
 \label{fig_rcomp_54}\includegraphics[width=60mm]{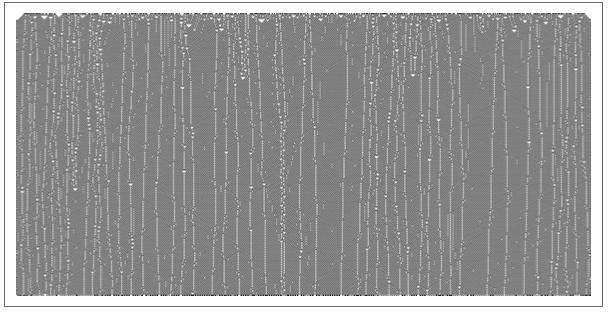}\\ 
 (b) Time evolution of ECA rule 54.
 \\[6pt]
   \label{fig_map_50_37}\includegraphics[width=100mm]{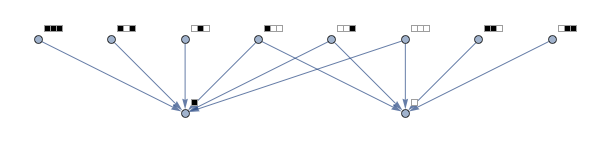} \\
(c) Non-causal rule mapping rule 50 $\circ$ 37. \\[6pt]\label{fig_rcomp_50_37b}\includegraphics[width=60mm]{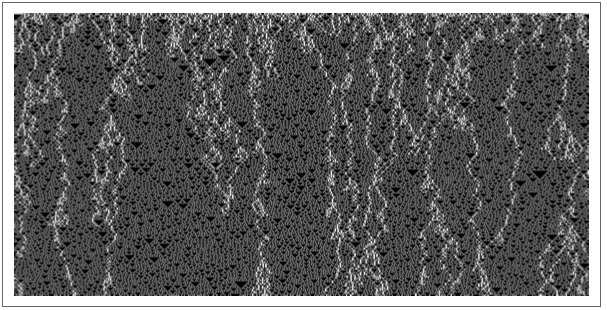} \\
(d) Time evolution of rule 50 $\circ$ 37. \\[6pt]
\end{tabular}
\caption{\label{fig_map_50_37b} Rule composition of ECA rules. (a) A network representation of the rule icon. (c) The rule icon is not causally separable and therefore the resulting composition is not an ECA rule but belongs to a larger CA rulespace. (b) Emulation of ECA rule 54. (d) Emulation of a non-ECA rule after Boolean composition of ECA rules. ECA is therefore not a closed space under composition.}
\end{figure}

Many rules can be composed from rule tuples not involving the composite rule itself. As in the case of PCA all ECA rules can be composed from other ECA rules. For example, for rule pairs there are $88^2=7744$ of which $7744-736=7008$ are not in the ECA rulespace. One could investigate these rules and determine to which Wolfram class they belong. For example, the rule pair (50, 37) (see Fig.~\ref{fig_map_50_37}) behaves like a Wolfram class 4 CA and seems potentially of high complexity. 

\begin{mydefinition}
A rule $R\in S$ in rulespace $S$ is $N$-prime if, and only if, it can only be simulated by itself or an equivalent rule under trivial symmetric transformations (see Supplementary Material) in $S$, i.e. no composition exists to simulate $R\in S$ up to the $N$-compositional iteration (see Algorithms in Subsection~\ref{samplingalg}) other than (possibly) a composition of $R$ itself.
\end{mydefinition}

\begin{mydefinition}
A rule $R^\prime \in S$ is $N$-composite if, and only if, it can be decomposed into a composition of other rules in $S$ non-equivalent to $R^\prime$ under trivial symmetry transformations (see Supplementary Material) up to an $N$-compositional iteration (see Algorithms in Subsection~\ref{samplingalg}).
\end{mydefinition}

\noindent It follows that prime and composite rules are disjoint subsets in the rulespace set of ECA.

\subsection{Order Parameters}

We use two order parameters that do not play any fundamental role in the main results yet offer a guide to the type of accepted general knowledge about generating rules and space-time evolution dynamics in ECA. Wolfram~\cite{nks} introduced a heuristic of behavioural class based on the typical behaviour of a CA for a random, equal density of non-zero (for binary) states/colours. Class 1 is the most simple (exhibiting the most trivial behaviour), followed by class 2 (converging to a uniform state); class 3 is random-looking and class 4 shows persistent structures and is thus considered the most complex.

\section{Proofs and Results}

We are interested in those rules which map back to ECA rules. We ask which is the minimal ECA rule set which produces all necessary tuples to compose all other ECA rules. One way to find such a set of `prime' rules is to create a graph having the rules as vertices and the edges created from the pairing of each tuple element to the composite rule. By looking at the vertex-in and vertex-out degrees one can eliminate the vertices which have a vertex-in degree $>0$ and a vertex-out degree $=0$. Taking the set of remaining vertices, i.e. rules, one can further eliminate vertices by exploring symmetries in the remaining graph. 

\begin{figure}[ht!]
\centering
\textbf{a}\hspace{5cm}\textbf{b}\\
  \includegraphics[width=45mm]{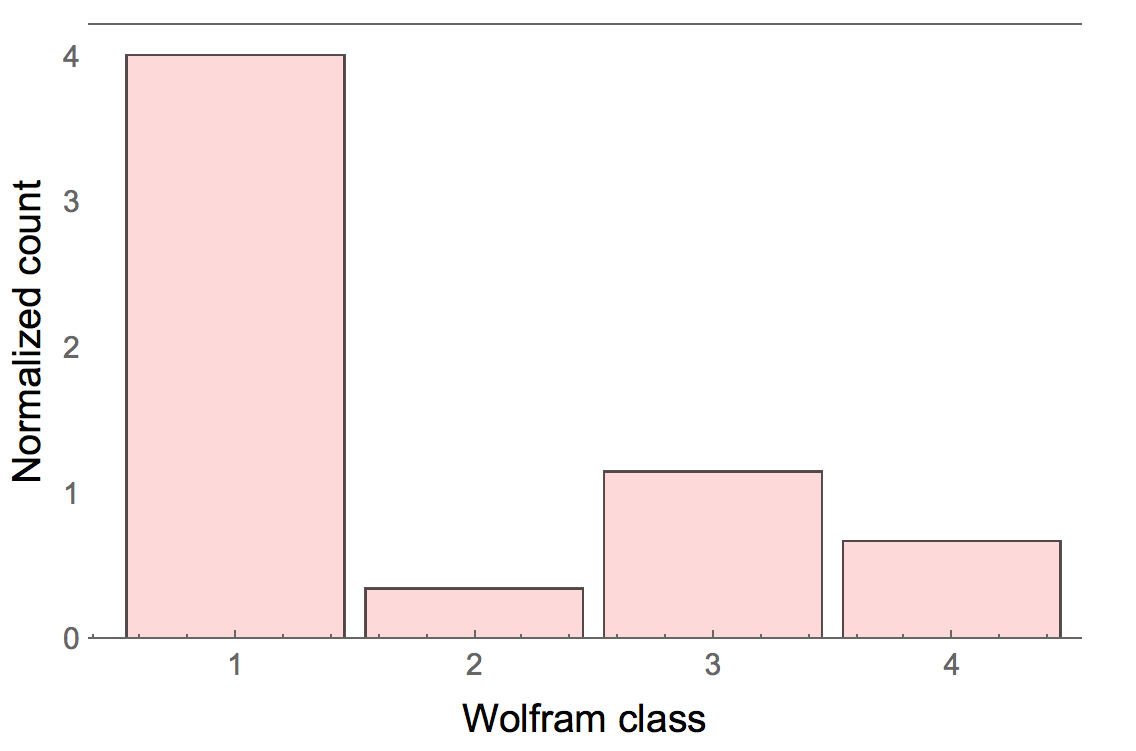}  \hspace{1cm} \includegraphics[width=45mm]{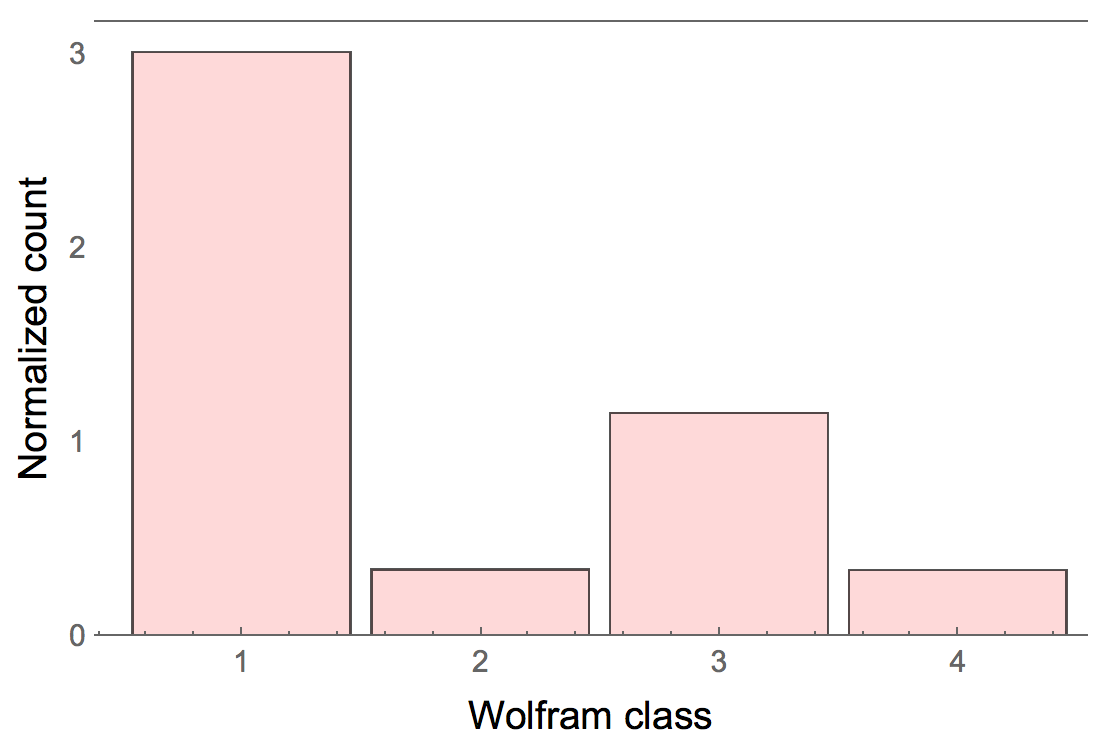}\\
  
  \medskip
  
 \textbf{c}\hspace{5cm}\textbf{d}\\
  \includegraphics[width=45mm]{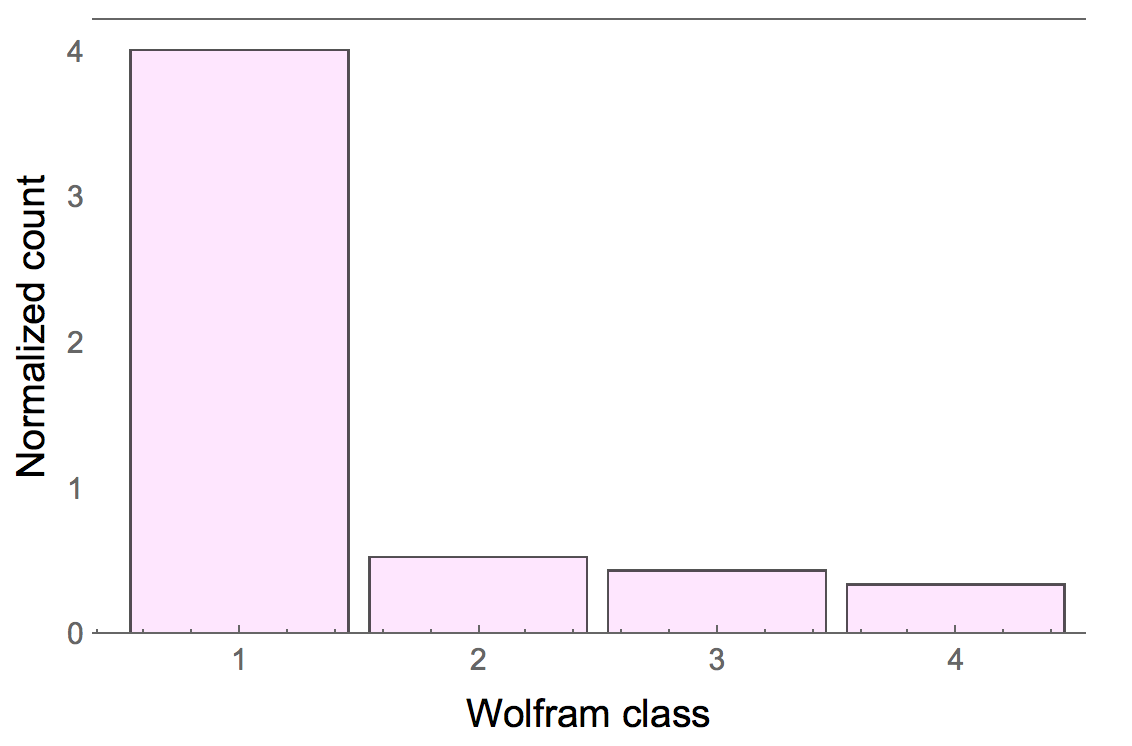}  \hspace{1cm} \includegraphics[width=45mm]{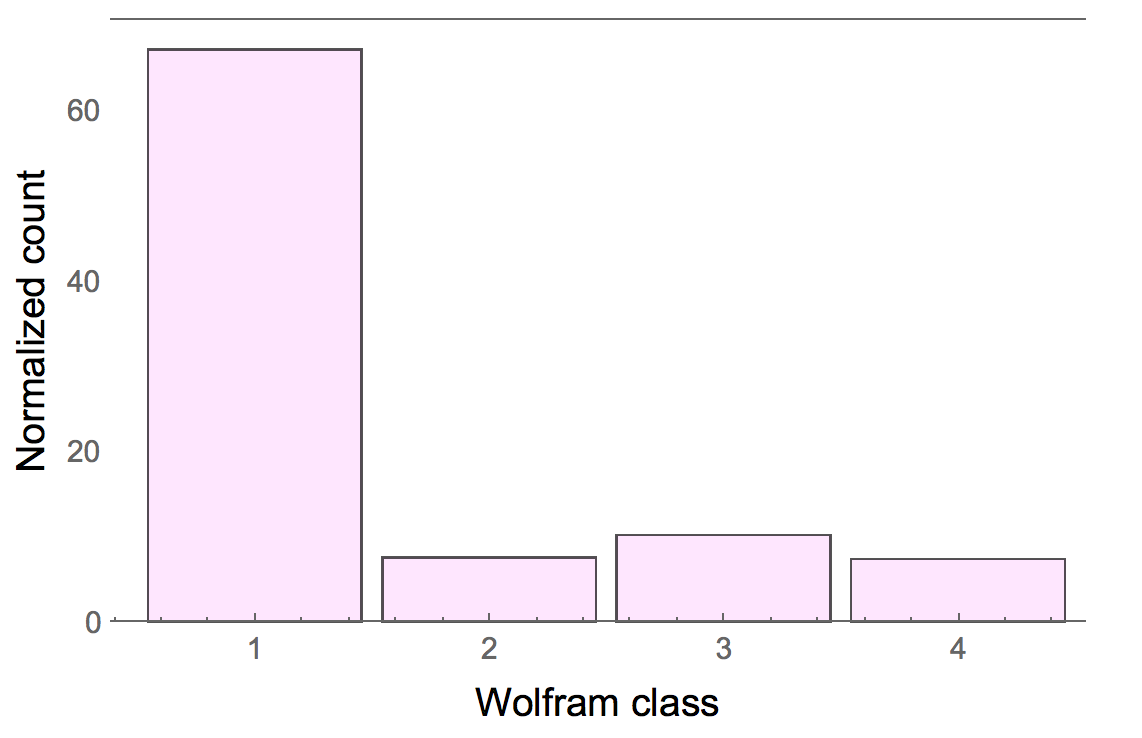}\\
   \medskip

 \textbf{e}\hspace{5cm}\textbf{f}\\
   \includegraphics[width=45mm]{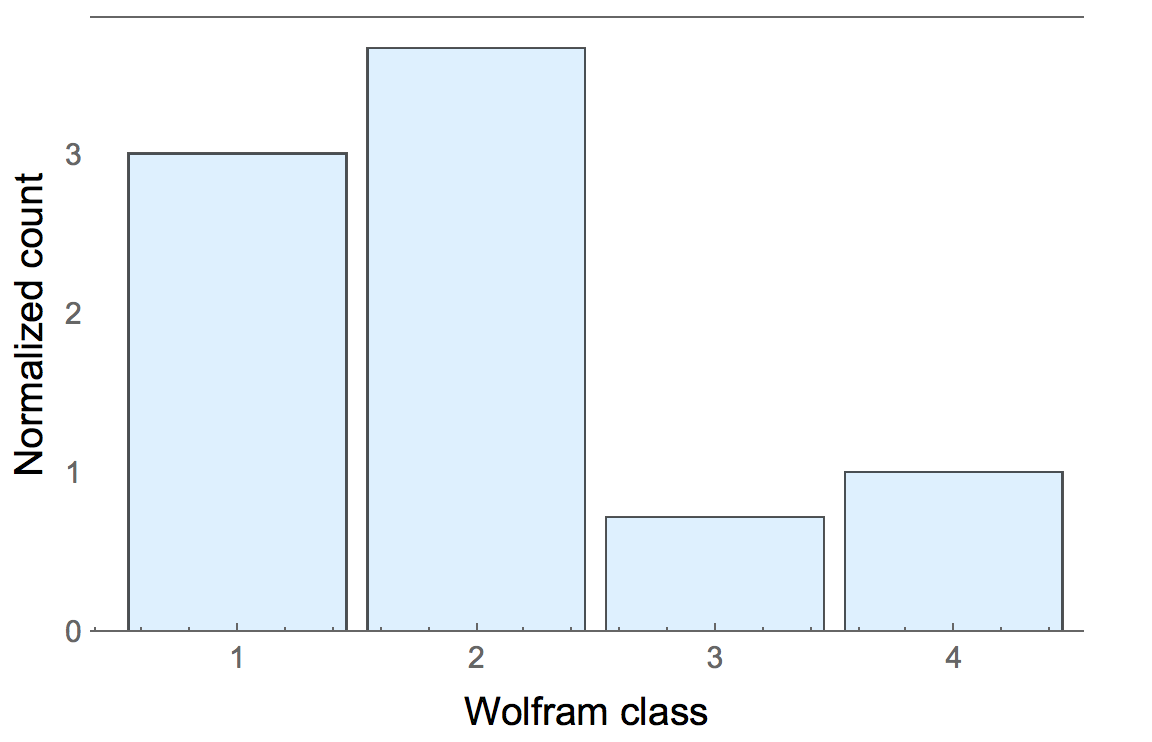}\hspace{1.2cm}\includegraphics[width=45mm]{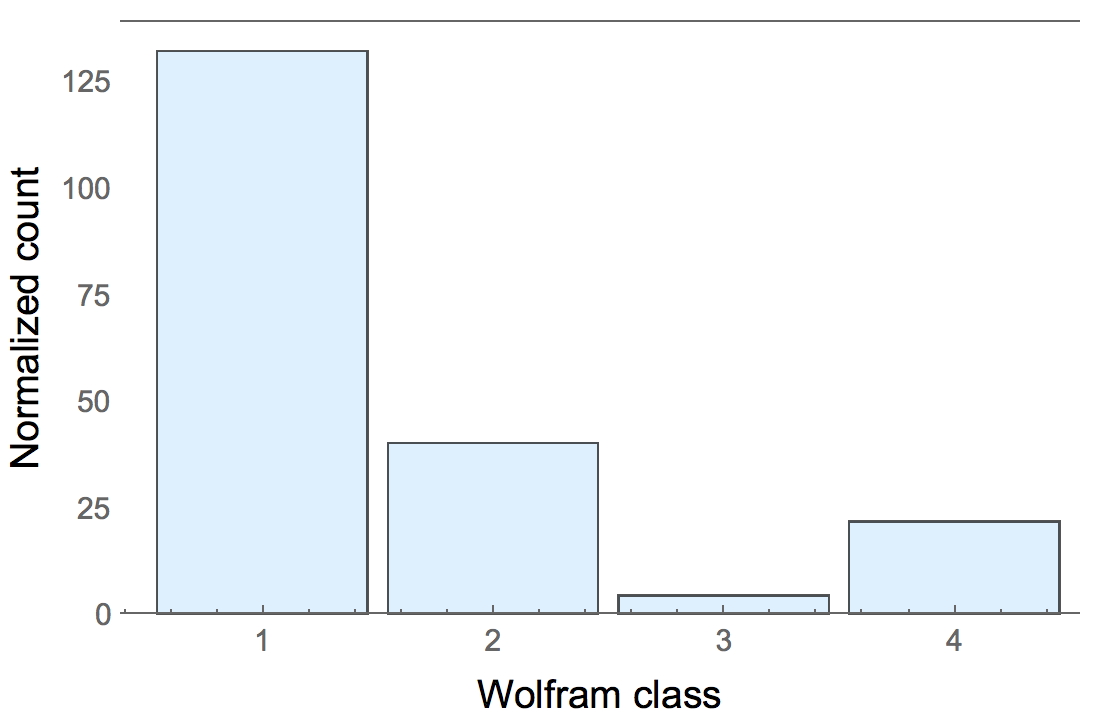}
\caption{\label{dists}(a,b) Distribution of primes by Wolfram class~\cite{nks} according to algorithms 1 (a) and 2 (b). (c,d) Distribution of vs composite rules by Wolfram class according to algorithms 1 (a) and 2 (b). Distributions of primes as used to generate all other rules under Boolean composition for algorithms 1 (e) and 2 (f). All bins are normalized by number of elements in each class for all 88 non-trivially symmetrical ECA rules.}
\end{figure}

\subsection{Rule Primality and Causal Decomposition}

Table.~\ref{primetable} shows the minimal ECA generating set with the prime rules and rule compositions for the composite rules.

\begin{myobservation}
None of the prime rules is of Wolfram class 4, i.e. all class 4 rules can be composed by prime rules of a lower class. In other words, all class 4 rules are composite.
\end{myobservation}

Fig.~\ref{dists} shows the distribution of Wolfram classes for prime and composite rules building the ECA space. Among the ECA prime rules used to generate all others, most belong to class 1 and 2, and these are the only 2 distributions for which sampling algorithms 1 and 2 produced the most different results. Yet in both cases rules 1 and 2 are the building blocks in the minimal ECA-generating sets. 

Table.~\ref{primetable} in the Supplementary Information provides all the compositions found and the breakdown of ECA prime and composite rules. It is common to find that rule permutations under compositions yield the same ECA rule. For example, rule 110 can be composed out of the prime tuple $170, 15, 118$. This is also true for some of the other permutations. Rule 54 can be composed out of the rule set (15, 108, 170) and all its permutations. However, this is not a rule, and no generating rule set (small or large) was found to be commutative or associative. 

However, the rule icons of prime rules most frequently used to generate all other ECA rules have similar, high non-zero state/colour density ($> 0.4$), with a Hamming distance of two to rule 110, yet they generate simple behaviour.

\subsection{Complexity of Prime and Composite Rules}

\begin{figure}
\centering
\textbf{a}\hspace{5.5cm}\textbf{b}\\
  \includegraphics[width=37mm]{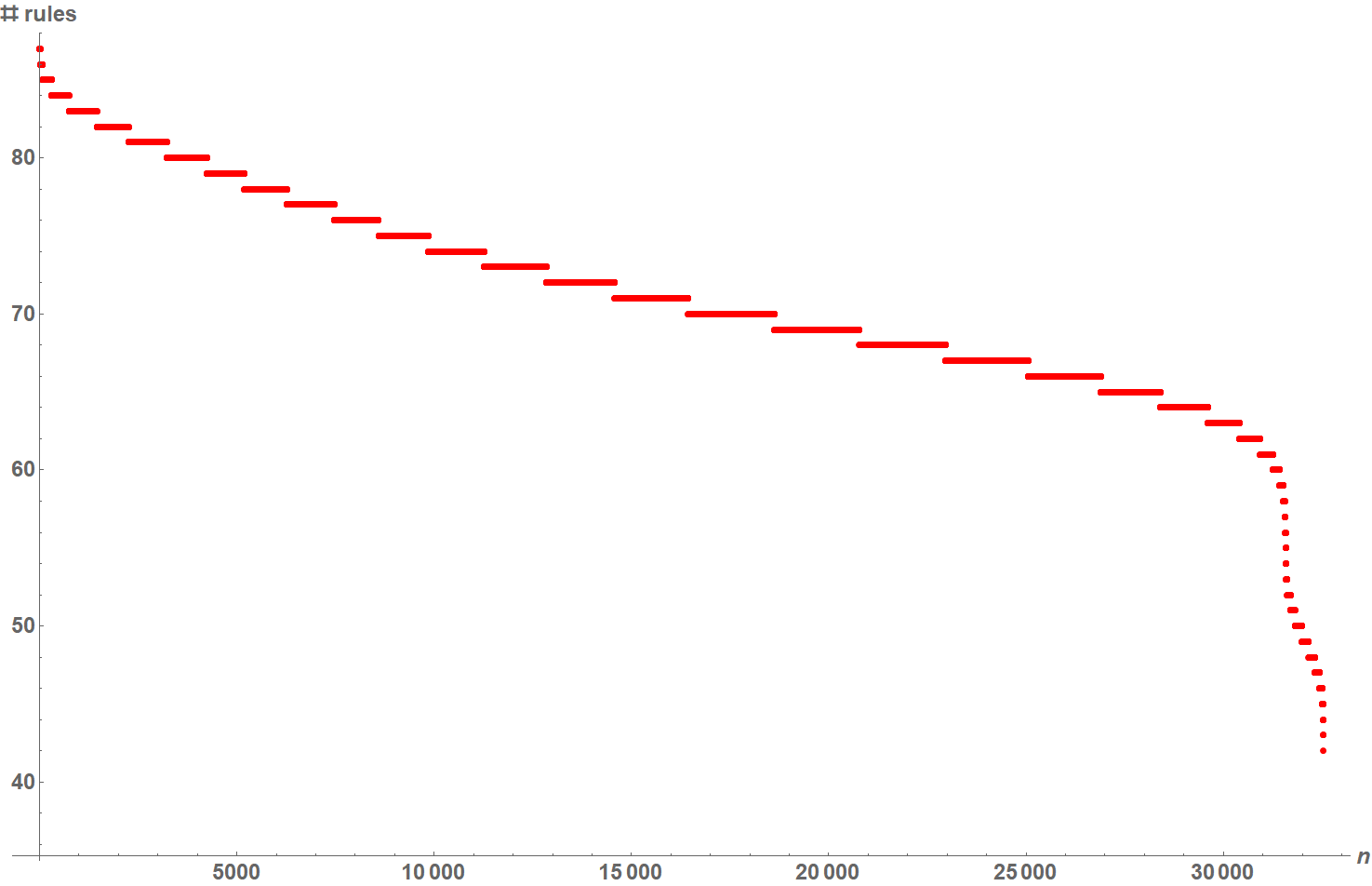}  \hspace{1cm} \includegraphics[width=37mm]{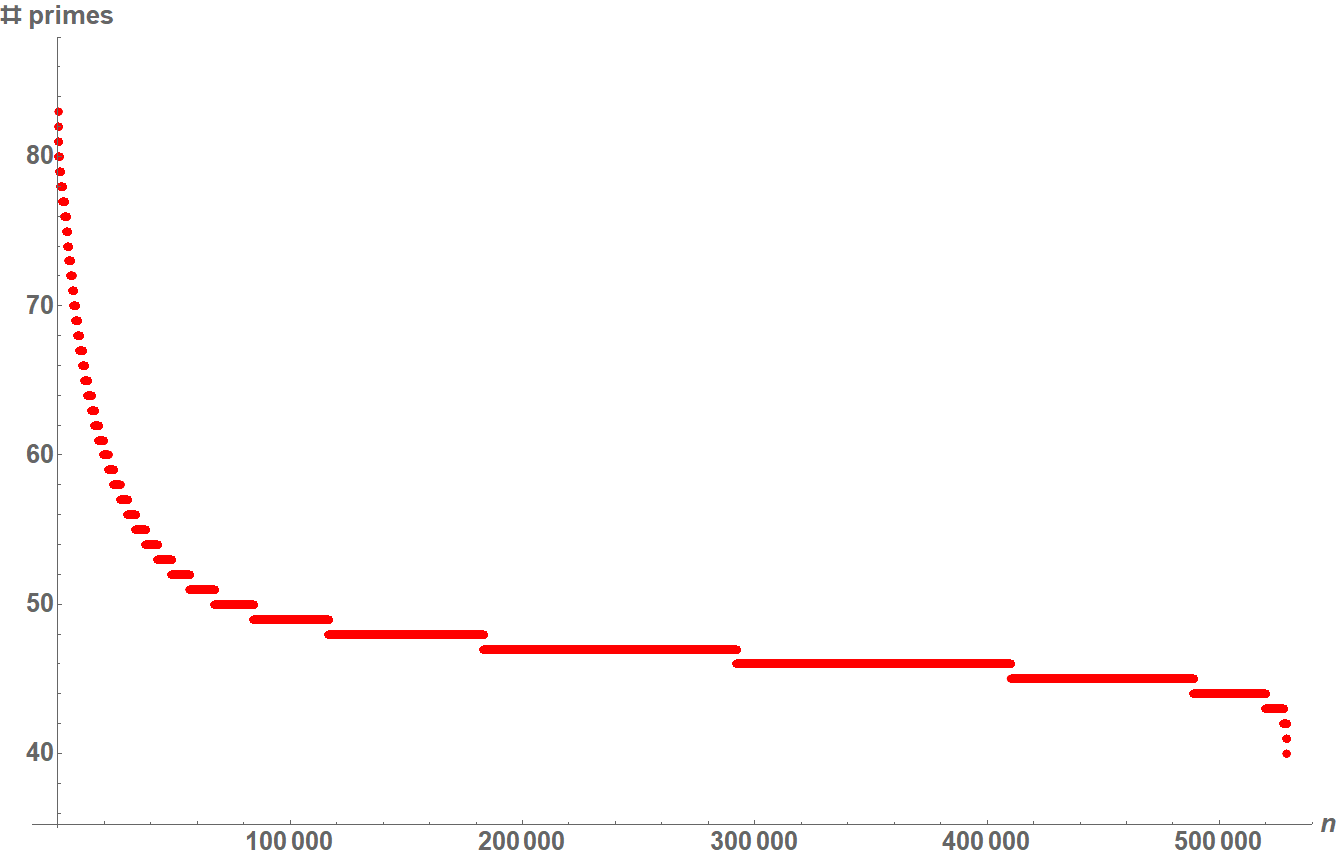}
  \medskip
  
\textbf{c}\\ 
  \includegraphics[width=83mm]{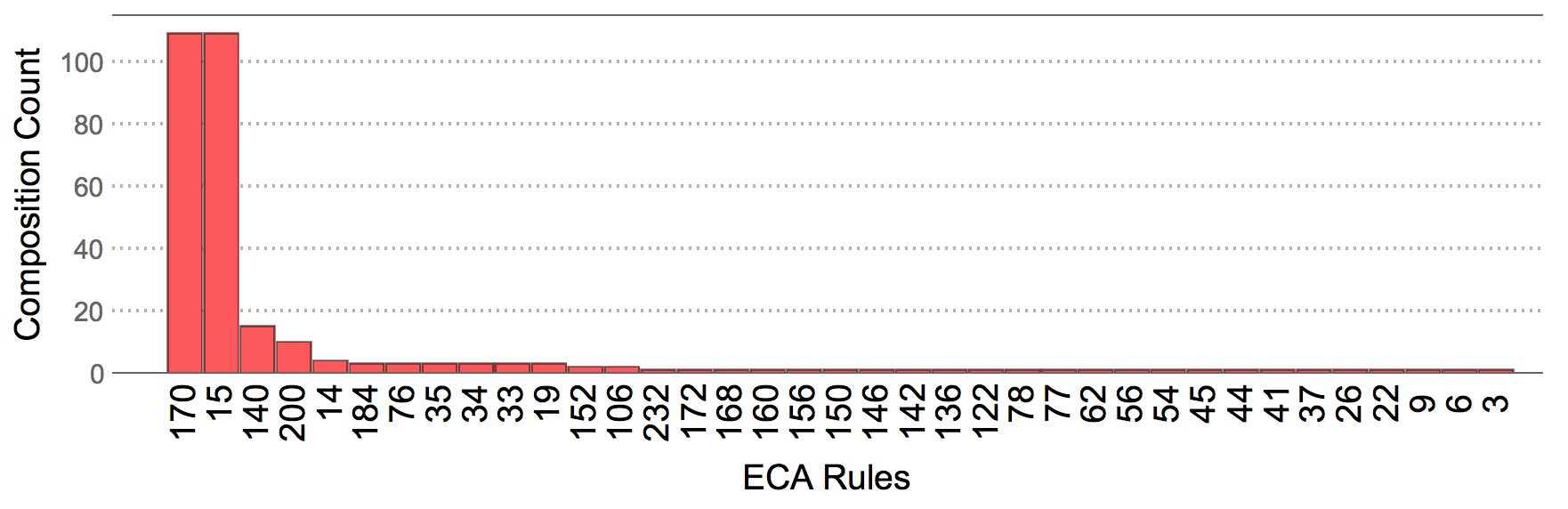}  \hspace{1cm}\\
    \medskip
    
\textbf{d}\\
  \includegraphics[width=83mm]{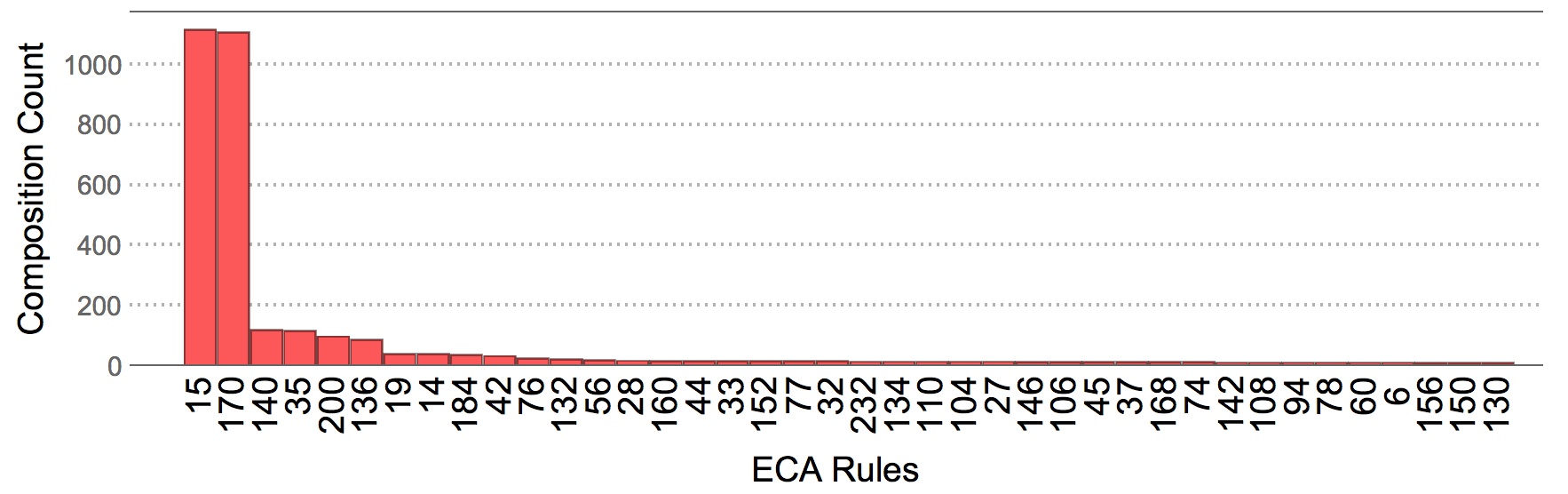}\\
    \medskip
    
    \textbf{e}\\
  \includegraphics[width=80mm]{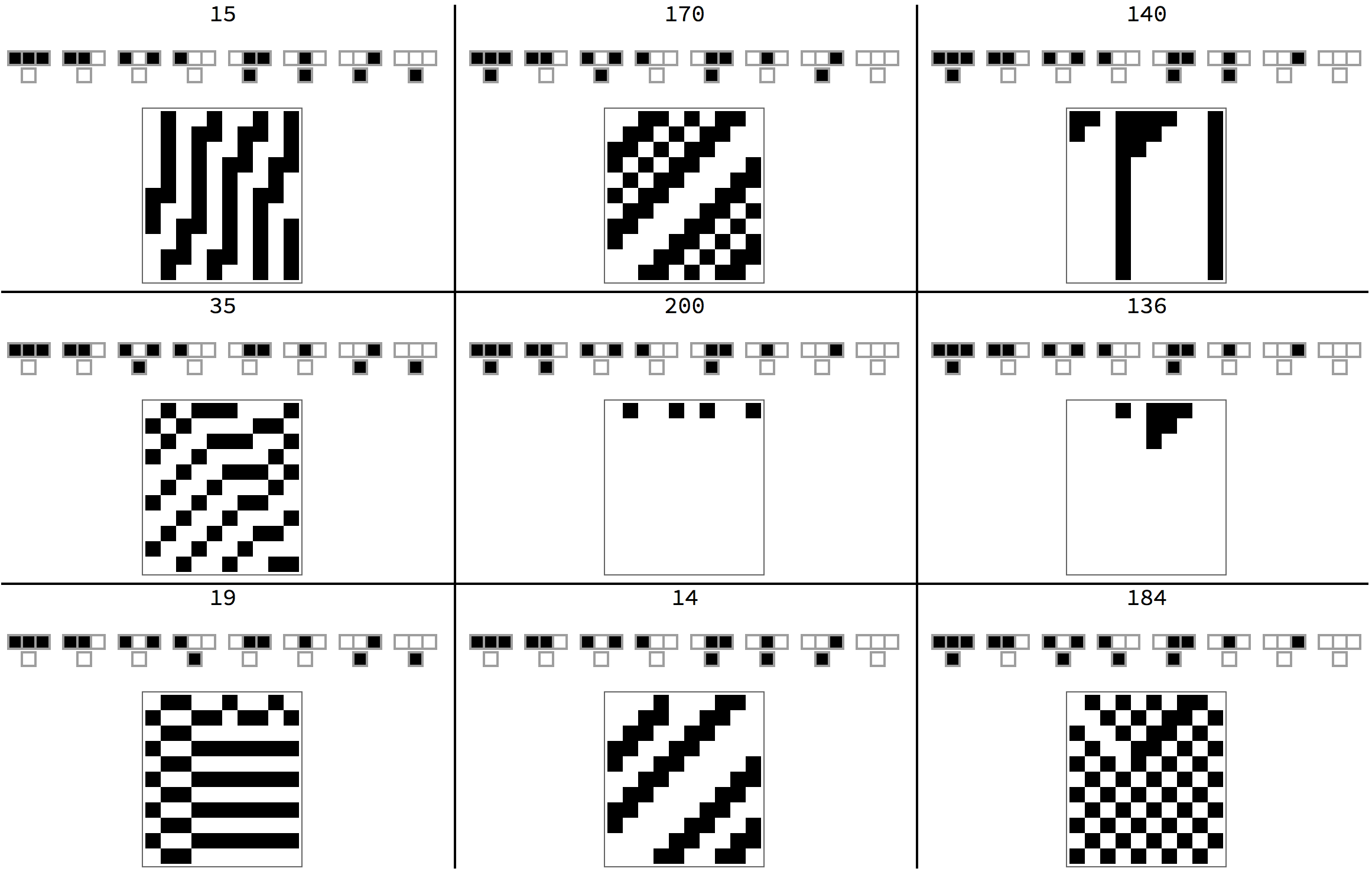}\\

\caption{\label{minimal}(a and b) Convergence to minimal sets, with the smallest of size $38$ ECA prime rules generating all other (256 or 88 non-equivalent) ECA rules for different seeds and two different algorithms with algorithm 2 (b) producing the smallest. (c) Top 40 most frequent prime rules able to produce all other ECA under Boolean composition according to algorithm 1 (d) very similar ranking of most frequent rules but with algorithm 2 among not only the minimal 38-element set but also all other 16 minimal sets of size at most 40. (e) ECA space-time evolutions of rules that are the building blocks of all other rules in the ECA space preceded by their rule icon running from random initial conditions for illustration purposes.}
\end{figure}

Fig.\ref{minimal}(c,d) shows the ECA prime rules with the highest frequency and their associated space-time evolutions, starting from a typical (i.e. random, 0.5 non-zero density) (Fig.\ref{minimal}(e)) among the `building blocks' able to generate the full ECA space (88 non-trivially symmetric rules or 256 rules counting all). They can be classified into two apparent main groups: (i) identity filters (e.g. rule 140, 136 and 200) able to partially `silence' or filter the communication of information from input to output and (ii) rules that transfer information diagonally or `shifters' (such as rules 170 and 14) at different speeds (e.g. slow, like rule 15, versus fast, like rules 14 and 184). The two most frequent ECA rules used to build all others are shifters that transfer information at different speeds with no collisions and no loss of information (rules 15 and 170).

The set of primes in the 38-rule minimal set able to produce all other ECA rules (88 non-equivalent and 256 under trivial symmetries) is: 0, 1, 2, 3, 5, 7, 11, 12, 13, 18, 19, 23, 24, 25, 27, 28, 29, 30,
34, 35, 38, 40, 42, 43, 46, 50, 51, 57, 58, 72, 73, 74, 94, 104, 105,
108, 110, 128, 130, 132, 134, 136, 138, 154, 162, 164, 178 and 204.

The composite rules in the minimal set are: 4, 6, 8, 9, 10, 14, 15, 22, 26, 32, 33, 36, 37, 41, 44, 45, 54, 56, 60, 62, 76, 77, 78, 90, 106, 122, 126, 140, 142, 146, 150, 152, 156,
160, 168, 170, 172, 184, 200 and 232.

The intersection between algorithm 1 and algorithm 2 is significant. Among all the 38 and 40 prime rule sets produced by the 2 algorithms, 27 are the same: 6, 9, 10, 14, 15, 22, 32, 37, 41, 60, 76, 77, 78, 90, 122, 126, 140, 142, 146, 152, 156, 160, 168, 170, 184, 200 and 232.

\subsection{Rule Decomposition}

\begin{figure}[ht!]
\centering
\textbf{a}\\
  \includegraphics[width=40mm]{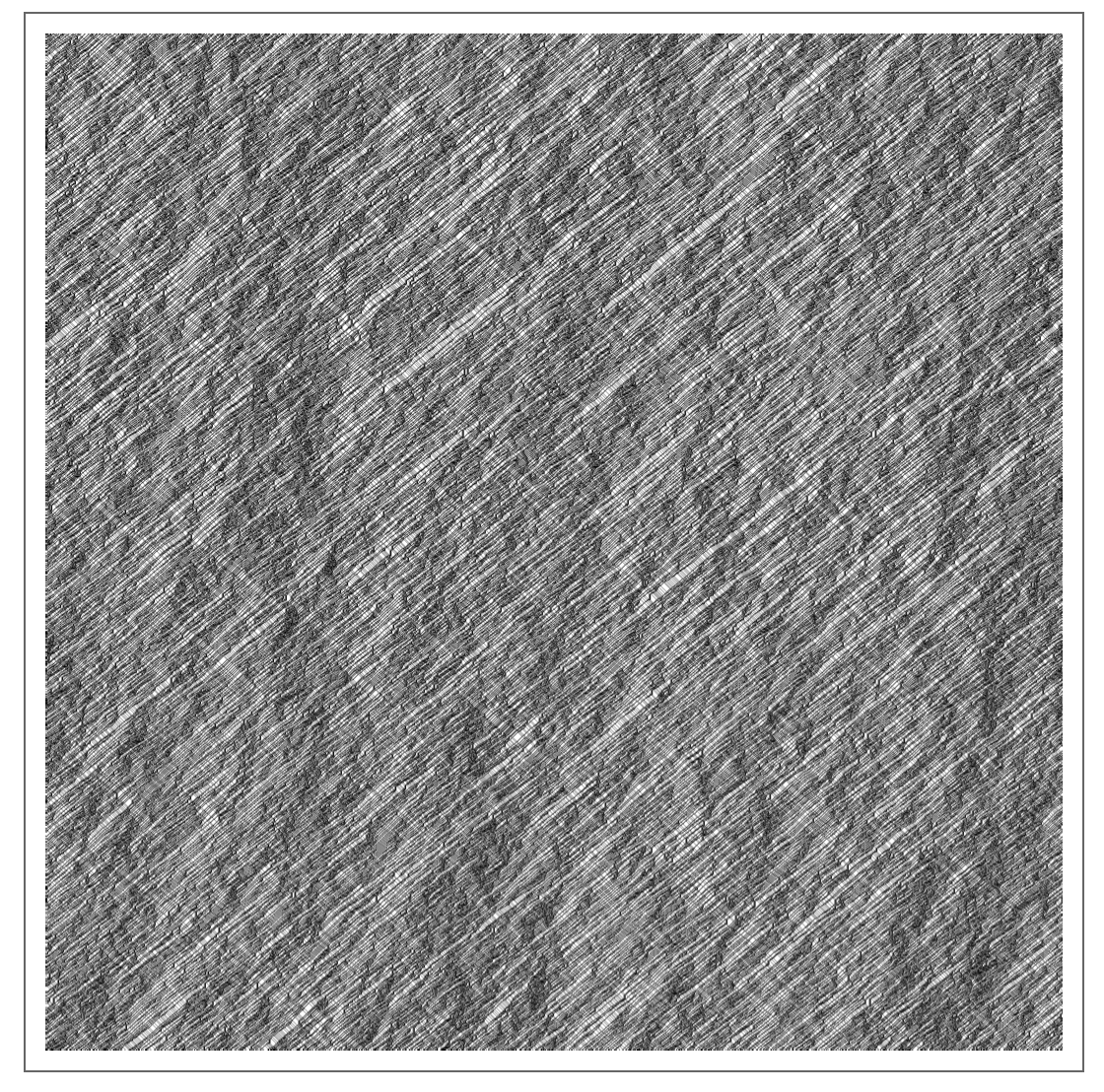}\\

\textbf{b}\hspace{3.2cm}\textbf{c}\hspace{3.2cm}\textbf{d}\\ 
  \includegraphics[width=25mm]{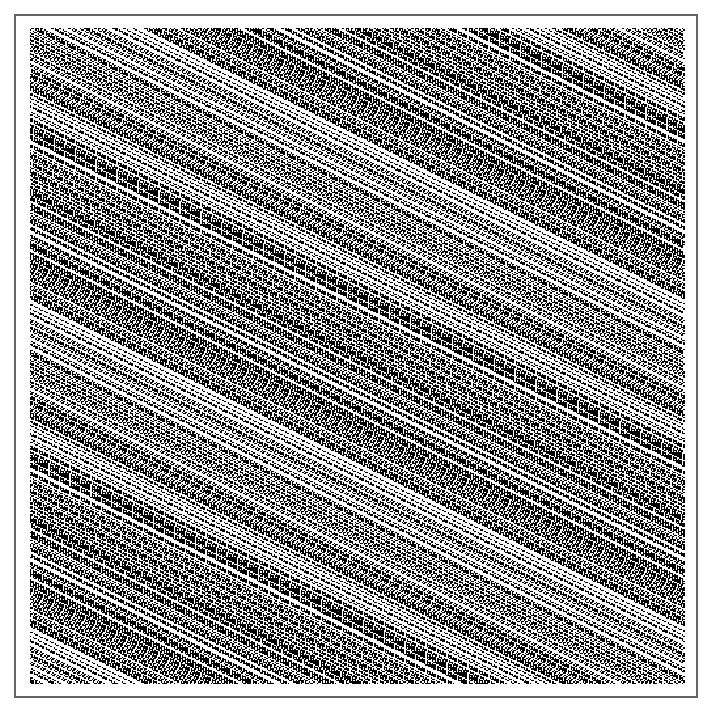}  \hspace{.5cm} \includegraphics[width=25mm]{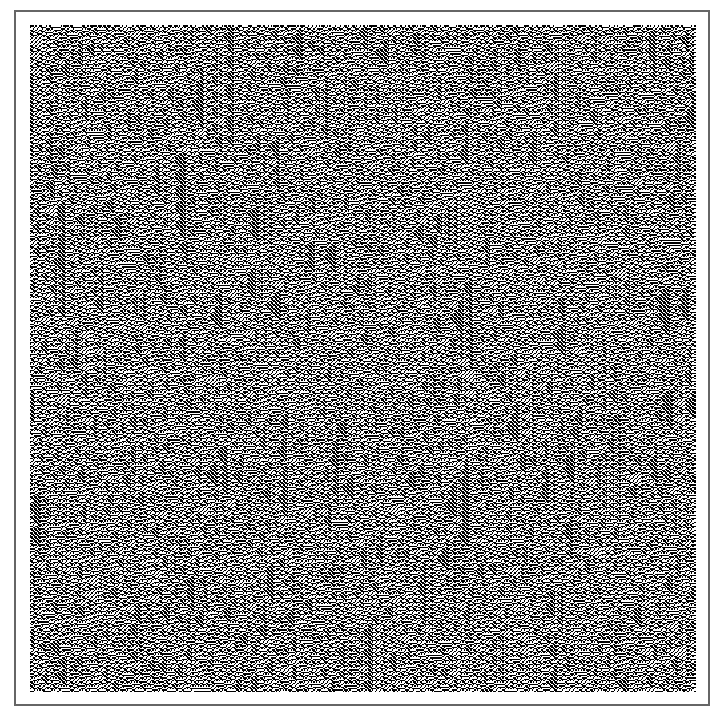} \hspace{.5cm} \includegraphics[width=25mm]{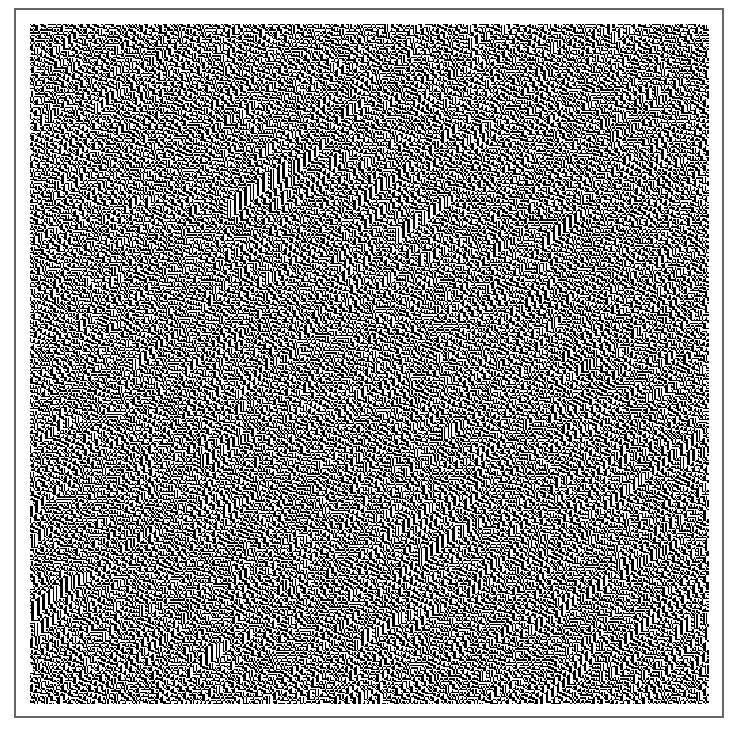}\\

  \textbf{e}\hspace{2.3cm}\textbf{f}\hspace{2.3cm}\textbf{g}\hspace{2.3cm}\textbf{h}\\
  \includegraphics[width=100mm]{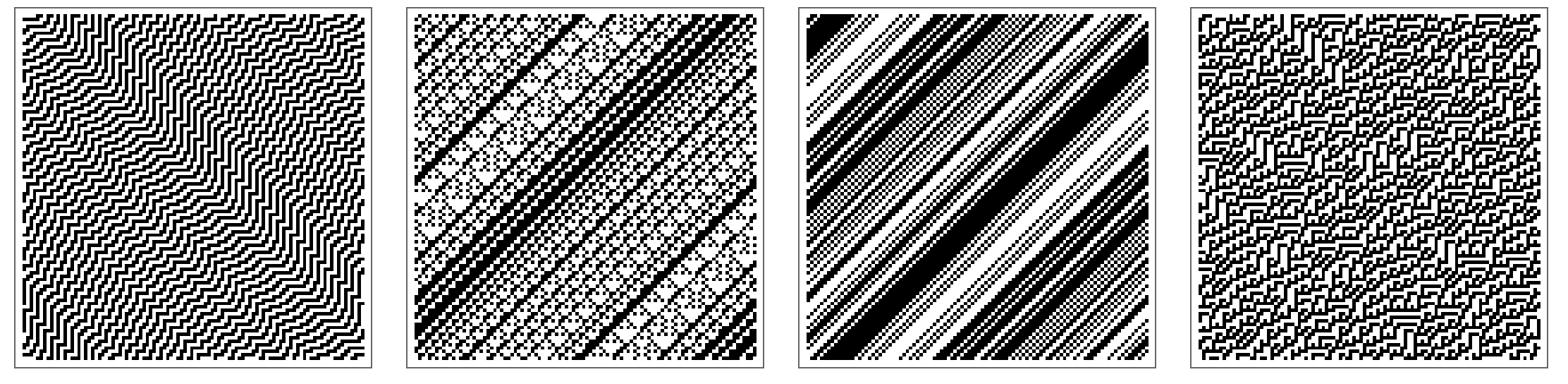} 
  
\caption{\label{decomposition}Causal decomposition: A complex CA (not in the ECA rule space) built by Boolean composition of (e,f,g,h) 4 simple ECA rules: 15 $\circ$ 154 $\circ$ 170 $\circ$ 45. (b,c,d) Pair compositions (15 $\circ$ 45), (15 $\circ$ 154), and (15 $\circ$ 154) of the same rules. Almost all permutations lead to the same behaviour. (f) and (g) build the diagonal `threads', (e) produces the fabric-like core and (h) introduces some random features that make (a) more realistic. All are run from random initial conditions for illustration purposes.}
\end{figure}

Fig.~\ref{decomposition} illustrates the way in which a set of simple ECA composed as building blocks in a Boolean circuit can produce rich behaviour, and how the different elements can causally explain the behaviour of the final system in a top-down and bottom-up way
through Boolean decomposition, fine- and coarse-graining.

\subsection{ECA Rule 110 Decomposition}

An analytic proof that ECA rule 110 can be composed out of prime ECA (170, 15, 118) is as follows:

\begin{figure}[ht!]
\centering
Emulation of rule 110: \hspace{2cm} Emulation of rule 54:
 \includegraphics[width=6.0cm,angle=0]{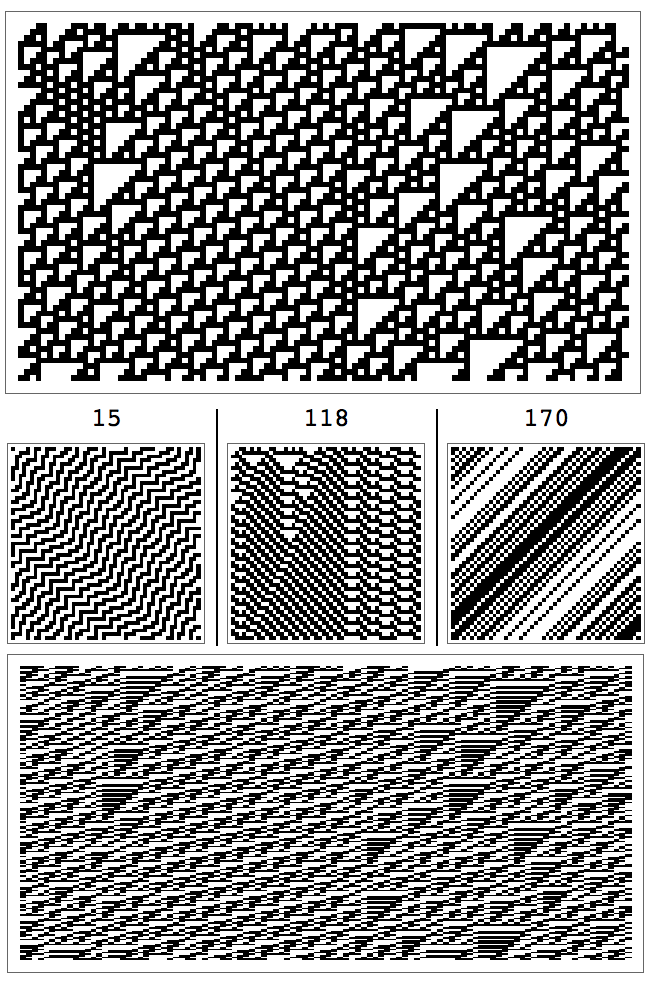}\includegraphics[width=6.0cm,angle=0]{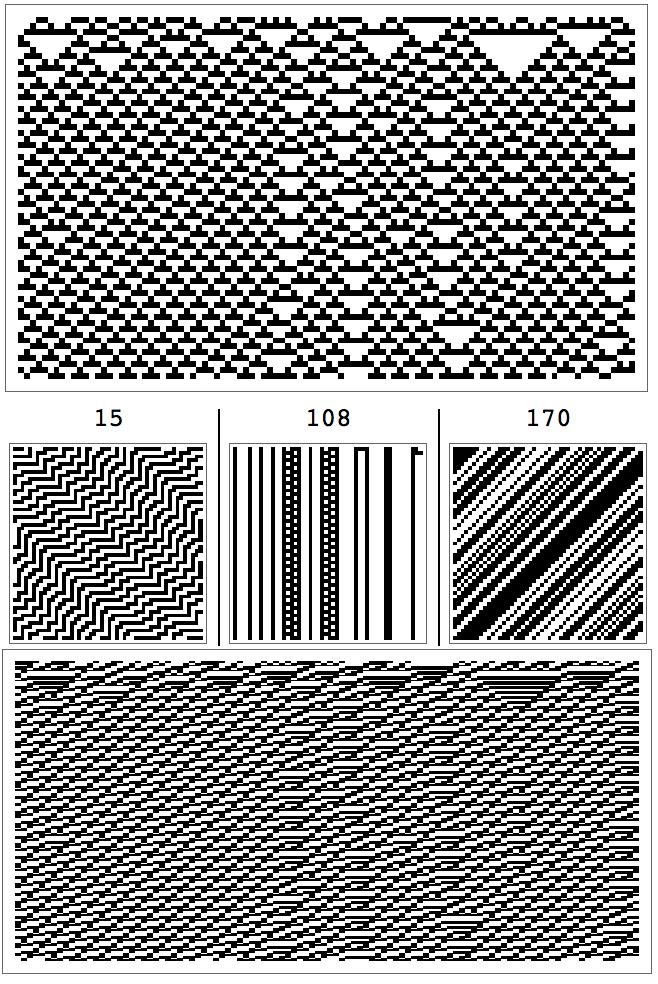}
\caption{\label{rule110emulation}Prime rule composition of ECA rule 110 (original rule on top) and emulation by composition of rules 15, 118 and 170 (middle) and of ECA rule 54 (same arrangement, but composition of rules 15, 108 and 170) after coarse-graining the stereographic version of the emulated rule (bottom).}
\end{figure}

\begin{mytheorem}
$\textrm{rule 110 } = \textrm{rule 51 } \circ \textrm{rule 118} = (\textrm{rule 170 } \circ \textrm{rule 15}) \circ \textrm{rule 118}$ 
\end{mytheorem}

The ECA rulespace clearly does not form a group because it is not closed under composition; no clear identity rule was found but identity candidates formed of prime rules were found. In particular, rules 15 and 170 are wild cards as prime rules able to perform bit shifts, and are used in almost every composition as they have the ability to target and shift a rule's bits.

\begin{proof}
First, we show that $\textrm{rule 51} = \textrm{rule 170} \circ \textrm{rule 15}$:

\bigskip

\noindent (a):  Given are the rules $\textrm{170}$: $(p,q,r) \mapsto r$ and $\textrm{15}$: $(p,q,r) \mapsto \neg p$

\begin{myenumerate}
\item Applying $\textrm{170}$: $(p,q,r) \mapsto r$ to the lattice, which shifts the lattice to the right: $p \rightarrow q, q \rightarrow r$, $r \rightarrow p$.
\item Applying now rule $\textrm{15}$: $(p,q,r) \mapsto \neg p$ we get: $q = \neg p \rightarrow q = \neg q \text{ (with 1.)} \text{ which is } \textrm{rule 51}: (p,q,r) \mapsto \neg q$.

\end{myenumerate}

Then, we show that  $\textrm{rule 110} = \textrm{rule 51} \circ \textrm{rule 118}$:

\bigskip

\noindent (b): Given that the rule 51: $(p,q,r) \mapsto \neg q$ and rule 118: $(p,q,r) \mapsto (p \vee q \vee r) \veebar (q \wedge r)$.

\begin{myenumerate}
\item Applying rule 51: $(p,q,r) \mapsto \neg q$ (rule 51) to the lattice, which negates the lattice.

\item Applying $(p,q,r) \mapsto (p \vee q \vee r) \veebar (q \wedge r)$ (rule 118) we get: $(\neg p \vee \neg q \vee \neg r) \veebar (\neg q \wedge \neg r)$

\item By De Morgan's Law: $\neg (p \wedge q \wedge r) \veebar \neg(q \vee r)$

\item Expanding xor: $(\neg (p \wedge q \wedge r) \wedge (q \vee r)) \vee ((p \wedge q \wedge r) \wedge \neg (q \vee r))$
\item =$((\neg p \vee \neg q \vee \neg r) \wedge (q \vee r)) \vee ((p \wedge q \wedge r) \wedge (\neg q \wedge \neg r))$
\item =$((\neg p \vee \neg q \vee \neg r) \wedge (q \vee r)) \vee ((p \wedge q \wedge r \wedge \neg q \wedge \neg r))$
\item =$((\neg p \vee \neg q \vee \neg r) \wedge (q \vee r)) \vee ((p \wedge r \wedge \neg r))$
\item =$((\neg p \vee \neg q \vee \neg r) \vee (p \wedge r \wedge \neg r) ) \wedge ((q \vee r) \vee (p \wedge r \wedge \neg r))$
\item =$((\neg p \vee \neg q \vee \neg r \vee p) \wedge (\neg p \vee \neg q \vee \neg r \vee r) \wedge (\neg p \vee \neg q \vee \neg r \vee \neg r)) \wedge ((q \vee r) \vee (p  \wedge r \wedge \neg r))$
\item =$(1 \wedge 1 \wedge (\neg p \vee \neg q \vee \neg r \vee \neg r)) \wedge ((q \vee r) \vee (p  \wedge r \wedge \neg r))$ 
\item =$(\neg p \vee \neg q \vee \neg r) \wedge ((q \vee r) \vee (p  \wedge r \wedge \neg r))$ \label{itm:interim}

\item $((q \vee r) \vee (p \wedge r \wedge \neg r))$ can be expanded as:
\item $(q \vee r \vee p) \wedge (q \vee r \vee r) \wedge (q \vee r \vee \neg r)$
\item =$(q \vee r \vee p) \wedge (q \vee r) \wedge 1$
\item =$(q \vee r) \wedge (1\wedge \vee p)$
\item =$(q \vee r)$
\item Substituting in (\ref{itm:interim}) one gets: $(\neg p \vee \neg q \vee \neg r) \wedge (q \vee r)$ \label{itm:interim1}
\item Starting now from rule 110: $(p,q,r) \mapsto (q \wedge \neg p) \vee (q \veebar r)$ by applying the definition of {\it xor}:
\item =$(q \wedge \neg p) \vee (q \wedge \neg r) \vee (\neg q \wedge r)$
\item =$(q \wedge \neg p \vee q) \wedge (q \wedge \neg p \vee \neg r) \vee (\neg q \wedge r)$
\item =$((q \wedge \neg p \vee q) \wedge (q \wedge \neg p \vee \neg r)) \vee \neg q) \wedge ((q \wedge \neg p \vee q) \wedge (q \wedge \neg p \vee \neg r)) \vee r) $ \label{itm:interim2}
\item The first part of (\ref{itm:interim2}) $((q \wedge \neg p \vee q) \wedge (q \wedge \neg p \vee \neg r)) \vee \neg q)$ can be expanded as:
\item =$((q \wedge \neg p) \vee q \vee \neg q) \wedge ((q \wedge \neg p) \vee \neg r \vee \neg q)$
\item =$(1 \wedge ((q \wedge \neg p) \vee \neg r \vee \neg q)$
\item =$((q \wedge \neg p) \vee \neg r \vee \neg q)$
\item =$((q \vee \neg r) \vee (\neg q \vee \neg r)) \vee \neg q$
\item =$(q \vee \neg r \vee \neg q) \vee (\neg p \vee \neg r \vee \neg q) $
\item =$1 \vee (\neg p \vee \neg r \vee \neg q) $
\item =$(\neg p \vee \neg r \vee \neg q) $ \label{itm:interim3}
\item The second part of (\ref{itm:interim2}) $((q \wedge \neg p \vee q) \wedge (q \wedge \neg p \vee \neg r)) \vee r)$ can be expanded as:
\item =$((q \wedge \neg p \vee q \vee r) \wedge (q \wedge \neg p \vee \neg r \vee r))$
\item =$((q \wedge \neg p \vee q \vee r) \wedge 1)$
\item =$((q \wedge \neg p) \vee q \vee r)$
\item =$(q \vee q \vee r) \wedge (\neg p \vee q \vee r)$
\item =$(q \vee r) \wedge (1 \vee \neg p)$
\item =$(q \vee r)$ \label{itm:interim4} 
\item Substituting in (\ref{itm:interim2}.) (\ref{itm:interim3}.) and (\ref{itm:interim4}.) one gets: $(\neg p \vee \neg q \vee \neg r) \wedge (q \vee r)$. \label{itm:interim5}
\item Since (\ref{itm:interim5}.) = (\ref{itm:interim1}.) this shows $\textrm{rule 110} = \textrm{rule 51} \circ \textrm{rule 118}$

\end{myenumerate}

\end{proof}
\begin{figure}
\centering
\begin{tabular}{cc}
  \label{fig_map_4color_110a}\includegraphics[height=50mm, width=50mm]{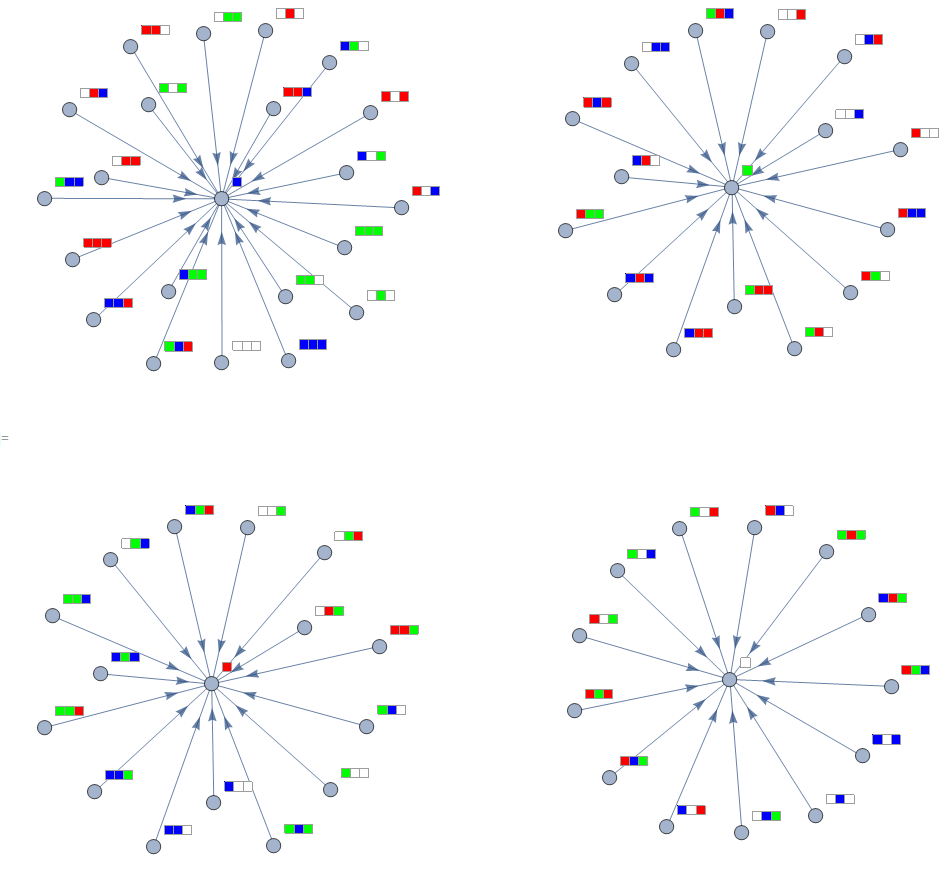} &   \label{fig_map_4color_110b}\includegraphics[height=50mm, width=50mm]{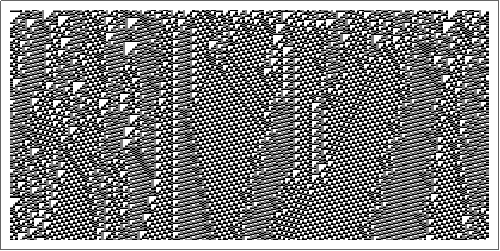} \\
(a) Rule mapping of 4-colour rule & (c) Space-time of 4-colour rule\\
170 $\circ$ 15 $\circ$ 118 & 170 $\circ$ 15 $\circ$118 with gray scale. \\[6pt]
 \label{fig_rcomp_110_2colorc}\includegraphics[height=50mm, width=50mm]{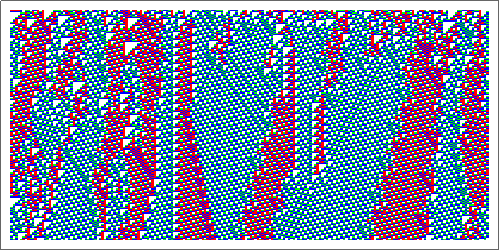} &   \label{fig_rcomp_110d}\includegraphics[height=50mm, width=50mm]{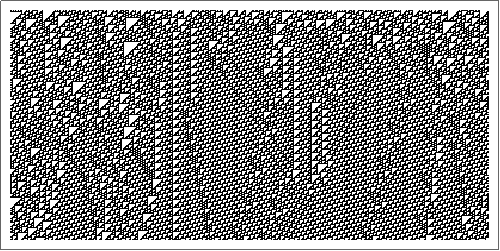} \\
(b) Spacetime of 4-colour rule & (d)  Space-time of rule \\
&170 $\circ$ 15 $\circ$ 118. 170 $\circ$ 15 $\circ$ 118.\\[6pt]
 \end{tabular}
\caption{\label{fig_4color_map_110} 4-colour rule equivalent to rule 170 $\circ$ 15 $\circ$ 118. Depicted are examples of space-time evolutions starting from random initial conditions (a) The rule icon is minimally separable and therefore the resulting composition is in the 4-colour CA rulespace. (b) Emulation of the 4-colour equivalent of rule 170 $\circ$ 15 $\circ$ 118 with colour mapping $\Box \rightarrow \Box \Box$, $\blacksquare \rightarrow \blacksquare \blacksquare$, $\mathcolour{red}{\blacksquare} \rightarrow \Box \blacksquare$ and $\mathcolour{green}{\blacksquare} \rightarrow \blacksquare \Box $. (c) Emulation of the 4-colour equivalent of rule 170 $\circ$ 15 $\circ$ 118 with colour re-mapping $\Box \rightarrow \Box$, $\blacksquare \rightarrow \blacksquare$, $\mathcolour{red}{\blacksquare} \rightarrow \mathcolour{gray}{\blacksquare}$ and $\mathcolour{green}{\blacksquare} \rightarrow \mathcolour{gray}{\blacksquare}$. The non-ECA 4-colour Turing-universal cellular automaton that simulates the rule space-time of the composition of ECA rules 170 $\circ$ 15 $\circ$ 118, that is, the Turing-universal ECA rule 110 after coarse-graining.}
\end{figure}

In a similar fashion, one can prove that the other Wolfram class 4 ECA rules 41, 54 and 106 are also composed of simpler prime rules.

\subsection{Multicolour CA Emulating ECA 110}

In order to find the CA in a higher rulespace that implements the Boolean composition of ECA emulating ECA rule 110, let's consider a block transformation of the form $\Box \Box \rightarrow \Box$, $\blacksquare \blacksquare \rightarrow \blacksquare$, $\Box \blacksquare \rightarrow \mathcolour{red}{\blacksquare}$ and $\blacksquare \Box \rightarrow \mathcolour{green}{\blacksquare}$ which maps all combinations of the 2-colour pairs with the 4 colours of the larger rulespace. In order to see if the rule icon of a CA generated by ECA rule composition is separable, one executes these steps:

\begin{enumerate}
\item Choose the de Bruijn sequence for alphabet  ${S_{A}}=\{{0,1,2,3}\}=\{{\Box,\blacksquare, \mathcolour{red}{\blacksquare},\mathcolour{green}{\blacksquare}}\}$ and sub-sequences of length $n=3$ as the initial condition.
\item Create 3-tuples representing the 4-colour rule tuples with range $r=1$
\item Apply the transformation  $\Box \rightarrow \Box \Box$, $\blacksquare \rightarrow \blacksquare \blacksquare$, $\mathcolour{red}{\blacksquare} \rightarrow \Box \blacksquare$ and $\mathcolour{green}{\blacksquare} \rightarrow \blacksquare \Box $ to each tuple.
\item Let the CA evolve each tuple 1 step for the chosen rule composition.
\item Apply back transformation  $\Box \Box \rightarrow \Box$, $\blacksquare \blacksquare \rightarrow \blacksquare$, $\Box \blacksquare \rightarrow \mathcolour{red}{\blacksquare}$ and $\blacksquare \Box \rightarrow \mathcolour{green}{\blacksquare}$ to each resulting output tuple.
\item Identify middle cell for each input tuple and pair it with the corresponding output 
tuple
\item Create a graph out off the resulting pairs.
\end{enumerate}

Fig.~\ref{fig_4color_map_110} illustrates the space-time evolution of this finer-grained CA capable of emulating ECA rule 110 after coarse-graining. The rule icon network for 170 $\circ$ 15$\circ$ 118 is separable in the 4-colour CA space and is thus a native CA belonging to the 2-colour rulespace with closest neighbour, the smallest rulespace in which such an automaton can exist. Fig.~\ref{fig_map_4colour_50_37} in the Supplementary Material illustrates another interesting example of causal decomposition.

\section{Conclusions}

We have introduced a notion of \textit{prime} and \textit{composite} rule that has allowed us to approach ECA from a group-theoretic point of view whose set composition can emulate all other Elementary Cellular Automata, suggesting minimal generating sets. While it was known that the set is not closed under Boolean composition and the emulations are not commutative, an exhaustive exploration of the emulating minimal sets had not been undertaken. The exploration allowed us to find some interesting emulations, including some Turing-universal compositions by emulation of ECA rule 110.

We found that two different sampling algorithms starting from different seeds reach and provide evidence that the smallest generating sets are close to 38 elements if not exactly 38, suggesting that simple rules are the building blocks of more complex rules in minimal generating sets. 

We have found features that appear to be essential in computation towards universality making a cellular automaton capable of emulating another (universal) cellular automata such as rule 110---and other complex rules---in the way in which these rules need to be composed with rules capable of transfer information horizontally at different rates. The new universality result in ECA is a composition of 2 and 3 rules but the actual Turing-universal cellular automaton is in a higher rulespace whose rule has been given in detail and is capable of emulating ECA rule 110 under coarse-graining.

We have introduced novel tools, concepts and methods to explore computation by algebraic/boolean rule composition, and methods for causal composition and decomposition. Our work suggests that novel model-based approaches to studying computational behaviour of computer programs can shed light on fundamental computational and causal processes underlying computing systems and provide a set of powerful tools to study general systems from a computational/informational perspective.

\newpage
\appendix

\section{Supplementary Material}

\subsection{Non-symmetric CA rules}

Local rules define the dynamical behaviour of CA. However, not all rules show essentially different dynamical behaviour. To focus on the number of rules in a rulespace which show essentially different dynamical properties, one can introduce the following symmetry transformations: \\

{\bf Reflection}: 
\begin{equation}
 f_r(x_1,x_2,...,x_n)=f(x_n,...,x_1,x_2)
\end{equation}

{\bf Conjugation}: 
\begin{equation}
 f_c(x_1,x_2,...,x_n)=q-1-f(q-1-x_1,q-1-x_2,...,q-1-x_n)
\end{equation}

{\bf Joint transformation}, i.e., conjugation and reflection:
\begin{equation}
 f_c\circ f_r (x_1,x_2,...,x_n)=q-1-f(q-1-x_n,...,q-1-x_1,q-1-x_2)
\end{equation}

Under these transformations two CA rules are equivalent, and they induce equivalence classes in the rulespace. Taking from each equivalence class a single representative (by convention the one with the smallest rule number), one gets a set which contains essentially different rules, i.e. rules which exhibit different global behaviour.

Let $G(\chi(f_r),\chi(f_c),\chi(f_cr))$ be a group under the operation $\circ$ acting on a set $X$ of all possible neighbourhood templates. Using the orbit counting theorem one can formulate the following theorem:

\begin{equation}
 \frac{1}{|G|}\sum_{g\in G}\chi(g)=\frac{\chi(f_I)+\chi(f_r)+\chi(f_c)+\chi(f_{cr})}{4}
\end{equation}
with $\chi(g)$ being the number of elements of $X$ fixed by $g$.

\begin{figure}
\footnotesize{
\begin{center}
\begin{tabular}{l|l|l} 
\textbf{Rule\textbf{:}Composition} & \textbf{Rule\textbf{:}Composition} & \textbf{Rule\textbf{:}Composition}\\
\hline
 0\textbf{:}3,32 & 35\textbf{:} & 108\textbf{:} \\
 \hline
 1\textbf{:}15,170,3,168,170,3,35,15,170 & 36\textbf{:} & 110\textbf{:}15,170,62 \\
 \hline
 2\textbf{:}170,3,35,15,170,14,15,170,35 & 37\textbf{:} & 122\textbf{:}15,170,94 \\
 \hline
 3\textbf{:} & 38\textbf{:} & 126\textbf{:}15,170,60,3,170 \\
 \hline
 4\textbf{:}15,170,32 & 40\textbf{:} & 128\textbf{:}3,35,170,15,170,3,168 \\
 \hline
 5\textbf{:}15,170,160 & 41\textbf{:}15,170,134 & 130\textbf{:}9,15,170 \\
 \hline
 6\textbf{:}15,170,40 & 42\textbf{:}14,15,170 & 132\textbf{:}15,170,33 \\
 \hline
 7\textbf{:}15,170,168 & 43\textbf{:} & 134\textbf{:} \\
 \hline
 8\textbf{:}3,35,170,14,15,170,35 & 44\textbf{:}15,170,38 & 136\textbf{:}3,170,3,35,15,170 \\
 \hline
 9\textbf{:} & 45\textbf{:}15,170,154 & 138\textbf{:}35,3,19,170,3,35,15,170 \\
 \hline
 10\textbf{:} & 46\textbf{:}14,15,170,35,25 & 140\textbf{:}15,170,35 \\
 \hline
 11\textbf{:}35,3,19 & 50\textbf{:}15,170,76 & 142\textbf{:}15,170,43 \\
 \hline
 12\textbf{:}14,15,170,35,14,15,170,35 & 51\textbf{:}15,170 & 146\textbf{:}15,170,73 \\
 \hline
 13\textbf{:}35,15,170,29 & 54\textbf{:}15,170,108 & 150\textbf{:}15,170,105 \\
 \hline
 14\textbf{:} & 56\textbf{:}15,170,28 & 152\textbf{:}15,170,25 \\
 \hline
 15\textbf{:} & 57\textbf{:} & 154\textbf{:} \\
 \hline
 18\textbf{:}170,3,35,15,170,76 & 58\textbf{:} & 156\textbf{:}15,170,57 \\
 \hline
 19\textbf{:}170,3,35,15,170,3,35,170 & 60\textbf{:} & 160\textbf{:} \\
 \hline
 22\textbf{:}15,170,104 & 62\textbf{:} & 162\textbf{:}15,170,35,15,170,29 \\
 \hline
 23\textbf{:}& 72\textbf{:}3,35,170,76 & 164\textbf{:}15,170,37 \\
 \hline
 24\textbf{:}25,14,15,170,35 & 73\textbf{:} & 168\textbf{:} \\
 \hline
 25\textbf{:} & 74\textbf{:}15,170,26 & 170\textbf{:} \\
 \hline
 26\textbf{:} & 76\textbf{:} & 172\textbf{:}15,170,27 \\
 \hline
 27\textbf{:} & 77\textbf{:} & 178\textbf{:}15,170,77 \\
 \hline
 28\textbf{:} & 78\textbf{:}15,170,58 & 184\textbf{:}15,170,29 \\
 \hline
 29\textbf{:} & 90\textbf{:} & 200\textbf{:}15,170,170,3,35,15,170 \\
 \hline
 30\textbf{:} & 94\textbf{:} & 204\textbf{:} \\
 \hline
 32\textbf{:} & 104\textbf{:} & 232\textbf{:}15,170,23 \\
 \hline
 33\textbf{:} & 105\textbf{:} &  \\
 \hline
 34\textbf{:}14,15,170,35,15,170 & 106\textbf{:}15,170,30 &  \\
\hline
 \end{tabular}
\end{center}
}
\caption{\label{primetable}A minimal generating set of the ECA rulespace. Rules that are not composite are prime rules; all others are composites. Each list on the right hand side is an ordered list of Boolean compositions for all 88 non-symmetric ECA rules. Rules are not interchangeable except in a few cases, and therefore ECA is a space that is neither closed nor commutative under composition.}
\end{figure}

\subsection{ECA Rulespace and Beyond}

The ECA rulespace contains 88 essentially different rules and 9 linear (additive) rules (0, 15, 51, 60, 90, 105, 150, 170, 204). The Wolfram classification groups the ECA rules as follows:

\begin{myitemize}
\item 8 Class 1 ECA rules (0, 8, 32, 40, 128, 136, 160, 168)
\item 65 Class 2 ECA rules (1, 2, 3, 4, 5, 6, 7, 9, 10, 11, 12, 13, 14, 15, 19, 23, 24, 25, 26, \
27, 28, 29, 33, 34, 35, 36, 37, 38, 42, 43, 44, 46, 50, 51, 56, 57, \
58, 62, 72, 73, 74, 76, 77, 78, 94, 104, 108, 130, 132, 134, 138, \
140, 142, 152, 154, 156, 162, 164, 170, 172, 178, 184, 200, 204, 232)
\item 11 Class 3 ECA rules (18, 22, 30, 45, 60, 90, 105, 122, 126, 146, 150)
\item 4 class 4 rules (41, 54, 106, 110)
\end{myitemize}

\begin{figure}
\centering
\begin{tabular}{cc}
  \label{fig_map_4colour_50_37}\includegraphics[height=50mm, width=50mm]{CA_50_37_4color_mapping.png} &   \label{fig_map_4colour_50_37b}\includegraphics[height=50mm, width=50mm]{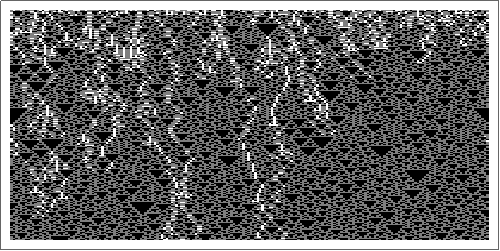} \\
(a) Rule mapping of 4-colour rule 50 $\circ$ 37. & (c) Space-time of 4-colour rule \\
& 50 $\circ$ 37 with gray scale. \\[6pt]
 \label{fig_rcomp_50_37_2colour}\includegraphics[height=50mm, width=50mm]{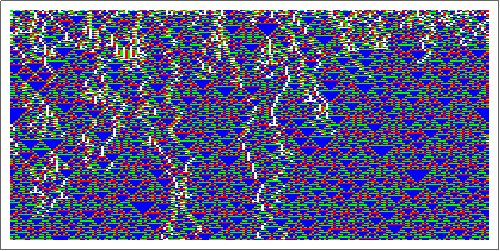} &   \label{fig_rcomp_50_37c}\includegraphics[height=50mm, width=50mm]{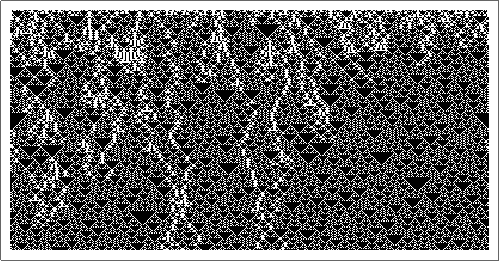} \\
(b) Space-time of 4-colour rule 50 $\circ$ 37. & (d) Space-time of rule 50 $\circ$ 37. \\[6pt]
 \end{tabular}
\caption{\label{fig_4colour_map_50_37}A 4-colour rule equivalent to rule 50 $\circ$ 37. Space-times start from random initial conditions for illustration purposes (a) The rule icon is separable and therefore the resulting composition can be represented in the larger 4-colour CA rulespace. (b) Emulation of the 4-colour equivalent of rule 50 $\circ$ 37 with colour mapping $\Box \rightarrow \Box \Box$, $\blacksquare \rightarrow \blacksquare \blacksquare$, $\mathcolour{red}{\blacksquare} \rightarrow \Box \blacksquare$ and $\mathcolour{green}{\blacksquare} \rightarrow \blacksquare \Box $. (c) Emulation of the 4-colour equivalent of rule 50 $\circ$ 37 with colour re-mapping $\Box \rightarrow \Box$, $\blacksquare \rightarrow \blacksquare$, $\mathcolour{red}{\blacksquare} \rightarrow \mathcolour{gray}{\blacksquare}$ and $\mathcolour{green}{\blacksquare} \rightarrow \mathcolour{gray}{\blacksquare}$. This `simulates' a black and white representation of the rulespace-time displayed. (d) Emulation of 2-colour rule 50 $\circ$ 37.}
\end{figure}

Since rule 50 $\circ$ 37 is not an ECA but a CA in a higher rulespace, we can check whether it also belongs to the 4-colour rulespace with range $r=1$ by finding the smallest separable icon network. 

\end{document}